\documentclass[a4paper,11pt]{article}
\pdfoutput=1
 \usepackage{jheppub} 
 \usepackage{graphicx}
\usepackage{amsmath}
 \usepackage{amssymb}
 \usepackage{slashed}
\usepackage{dcolumn}
\usepackage{bm}
\usepackage{color}
\usepackage{multirow}

\allowdisplaybreaks

 \newcommand{\lsim}{{\;\raise0.3ex\hbox{$<$\kern-0.75em\raise-1.1ex\hbox{$\sim$}}\;}}
\newcommand{\gsim}{{\;\raise0.3ex\hbox{$>$\kern-0.75em\raise-1.1ex\hbox{$\sim$}}\;}}
\def\bea{\begin{eqnarray}}
\def\eea{\end{eqnarray}}
\def\bec{\begin{center}}
\def\ec{\end{center}}

\def\beq{\begin{equation}}
\def\eeq{\end{equation}}

\def\bea{\begin{eqnarray}}
\def\eea{\end{eqnarray}}
\def\beq#1\eeq{\begin{align}#1\end{align}}
\def\beqnn#1\eeq{\begin{align*}#1\end{align*}}
\def\ba{\begin{array}}
\def\ea{\end{array}}
\def\bc{\begin{center}}
\def\ec{\end{center}}

\preprint{CTPU-PTC-19-30,\, KIAS-P19059}

\title{Axion-photon-dark photon oscillation and its implication for 21\,cm observation}
 
\author{
    Kiwoon Choi$^{a}$\footnote{Electronic address: kchoi@ibs.re.kr},
    Hyeonseok Seong$^{b,a}$\footnote{Electronic address: sbravos@kaist.ac.kr},
    Seokhoon Yun$^{c}$\footnote{Electronic address: SeokhoonYun@kias.re.kr}}
     
\affiliation{
 $^a$Center for Theoretical Physics of the Universe,  Institute for Basic Science, Daejeon 34126,\\
  South Korea \\
  $^b$Department of Physics, KAIST, Daejeon 34141, South Korea \\
 $^c$Korea Institute for Advanced Study, Seoul 02455,  South Korea  
 }

\abstract{
We examine the resonant conversion of axion-like particle (ALP) or dark photon to the electromagnetic photon in the early Universe, which takes place due to the ALP-photon-dark photon oscillations in background dark photon gauge fields. It is noted that the corresponding conversion probability can have an unusual spectral feature which allows strong conversion at low frequency domain, but has negligible conversion at high frequencies above certain critical frequency which is  determined by the ALP coupling to dark photon and the strength of background dark photon gauge field. We apply this scheme 
to heat up  the 21\,cm photons without affecting the Cosmic Microwave Background, which can explain
the tentative  absorption signal of 21\,cm photons detected recently by the EDGES experiment.     
}

\vspace{4cm}

\begin{document} 
\maketitle
\flushbottom

\section{Introduction}
\label{sec:Intro}

In spite of the great success of the Standard Model (SM) of particle physics, there are many reasons to contemplate new physics  beyond the Standard Model (BSM).
Although it needs to be confirmed later, the recent EDGES data \cite{Bowman:2018yin} of the Rayleigh-Jeans tail of the Cosmic Microwave Background (CMB) may provide another indication of BSM physics as it can be interpreted as an anomalously strong absorption signal of the 21\,cm photons.
There can be two approaches to explain the EDGES anomaly, either cool down baryons to lower the spin temperature \cite{Barkana:2018lgd} or heat up the 21\,cm photons. 
For the latter approach, an efficient way is to utilize resonant conversion of dark radiations (DR)  to 21\,cm photons in the early Universe, which occurs at the redshift in the range  $20 < z < 1700$~\cite{Pospelov:2018kdh,Moroi:2018vci}. 
Such scenario involves an ultra-light DR with a mass satisfying the resonance condition $m_{\rm DR}=m_\gamma(z)= {\cal O}(10^{-14}-10^{-9})\, {\rm eV}$ for $20<z<1700$, where $m_\gamma(z)$ is the effective photon mass in the early Universe, as well as an appropriate coupling of DR to induce the necessary conversion to photons. 
There are two appealing candidates for such DR,  an axion-like particle (ALP) and  a dark photon \cite{Pospelov:2018kdh,Moroi:2018vci}.

Nonetheless, the scheme to heat up  21\,cm photons  is facing with many observational constraints and also theoretical concerns on its naturalness. One of the major constraints comes from
 the distortion of CMB due to the conversion of CMB to DR  \cite{Mirizzi:2009iz,Mirizzi:2009nq}.   
Yet its significance severely depends on the spectral feature of the conversion probability, which is often determined by the scaling property of the underlying coupling.
 For instance, for the conversion of dark photons to 21\,cm photons  \cite{Pospelov:2018kdh}, the underlying coupling is the mass-dimension $d=4$ kinetic mixing \bea\frac{1}{2}\varepsilon F_{\mu\nu} X^{\mu\nu},\eea where $F^{\mu\nu}$ and $X^{\mu\nu}$ are the field strengths of the electromagnetic photon $\gamma$ and the dark photon $\gamma^\prime$, respectively.
One then finds the resonant conversion probability $P_{\gamma\leftrightarrow \gamma^\prime}\propto \varepsilon^2/\omega$ \cite{Pospelov:2018kdh,Mirizzi:2009iz}, where the dependence on the photon frequency $\omega$ originates from  relativistic kinematics.   As $\omega_{\rm CMB}\sim 10^3 \,\omega_{21}$, the conversion  at the CMB frequency $\omega_{\rm CMB}$
is suppressed relative to the conversion at the 21\,cm frequency $\omega_{21}$. This makes it possible that  sizable  parameter region can avoid a dangerous distortion of CMB.
However this scenario needs a symmetry breaking sector to generate a tiny dark photon mass $m_{\gamma^\prime}={\cal O}(10^{-14}-10^{-9})\, {\rm eV}$, which
may cause a
severe naturalness problem or require an uncomfortably low cutoff scale of the model as was discussed recently in \cite{Reece:2018zvv}. It requires also a small kinetic mixing $\varepsilon < 10^{-5}$ \cite{Pospelov:2018kdh}, which may cause another fine tuning problem in the UV completion of the model.    

For the scenario utilizing the resonant conversion of ALP to photons in background magnetic field \cite{Moroi:2018vci}, 
the required small ALP mass  $m_a ={\cal O}(10^{-14}-10^{-9})\, {\rm eV}$ can be achieved 
 by a tiny non-perturbative  breaking of the ALP shift symmetry $U(1)_a: a\rightarrow a+{\rm constant}$
without causing a naturalness problem\footnote{Although quantum gravity arguments suggest that the ALP shift symmetry $U(1)_a$
can not be an exact symmetry \cite{Harlow:2018tng}, it is yet a plausible possibility  that $U(1)_a$ is preserved in perturbation theory. In such case, $U(1)_a$ is broken mostly by non-perturbative effects which can be exponentially suppressed in certain region of the moduli space in the underlying theory.}. 
  The main difficulty of this scenario originates from the observational constraints on 
the underlying ALP coupling  \bea
\frac{1}{4}g_{a\gamma\gamma}a F^{\mu\nu} \tilde F_{\mu\nu},\eea where $\tilde F_{\mu\nu}=\frac{1}{2}\epsilon_{\mu\nu\rho\sigma} F^{\rho\sigma}$ is the dual electromagnetic field strength.
As this coupling has a  mass-dimension $d=5$, the conversion is more efficient at higher frequency as $P_{\gamma\leftrightarrow a}\propto g_{a\gamma\gamma}^2\omega$.  Then
the conversion at $\omega_{\rm CMB}$ is much stronger than the conversion at $\omega_{21}$, 
 making the constraint from the CMB distortion  quite severe.
 There exist also other constraints to be taken, for instance an upper bound on  $g_{a\gamma\gamma}$ from the absence of $\gamma$-ray burst associated with  SN1987A \cite{Payez:2014xsa},
and also an upper bound on the primordial background magnetic field, $B_0\lesssim 0.1$ nG, to avoid an overheating of baryons which would wash away the EDGES signal \cite{Minoda:2018gxj}. 
As is presented in  Appendix \ref{app:ALPrevision}, taking these
 constraints together, only a tiny parameter region of $(m_a, g_{a\gamma\gamma})$ can provide a viable explanation for the EDGES anomaly  even when one assumes the most optimistic spectrum of
ALP dark radiation and also a primordial background magnetic field $B_0\sim 0.1$ nG which is close to its upper bound.

In this paper, we wish to explore an alternative scheme to explain the EDGES anomaly. Our scheme involves both an ALP with $m_a\lesssim 10^{-9}\, {\rm eV}$ and a massless dark photon. It is utilizing again the resonant conversion of ALP or dark photon to 21\,cm photons, but based on the photon-ALP-dark photon oscillation in background dark photon gauge field, which is induced by the 
ALP couplings
\bea
\frac{1}{2}g_{a\gamma \gamma^\prime}aF_{\mu\nu}\tilde{X}^{\mu\nu} +\frac{1}{4}g_{a\gamma^\prime \gamma^\prime}aX_{\mu\nu}\tilde{X}^{\mu\nu}.
\label{eq:alp-coupling}
\eea 
As the dark photon $X_\mu$ is exactly massless, 
we don't need an additional sector to break the dark photon $U(1)_X$ gauge symmetry.  
 A key ingredient of our scheme is a nonzero primordial background dark photon gauge field strength $\langle X_{\mu\nu}\rangle$, which can be easily generated in the early Universe as was demonstrated in
\cite{Choi:2018dqr}.  In the presence of $\langle X_{\mu\nu}\rangle\neq 0$, the ALP coupling $g_{a\gamma \gamma^\prime}$ 
induces a mixing between $\gamma$ and ALP, while
$g_{a\gamma^\prime \gamma^\prime}$ induces a mixing between another pair, $\gamma^\prime$ and ALP. As a consequence, the two ALP couplings in (\ref{eq:alp-coupling}) result in oscillations among the three different particle states $\gamma$, $\gamma^\prime$ and ALP in background  $\langle X_{\mu\nu}\rangle$.

As for the spectral dependence of the conversion probability $P_{\gamma\leftrightarrow a,\gamma^\prime}$ in our scheme,   
it reveals an unusually interesting feature.  The conversion probability  
can be large enough, even close to unity,  over a certain frequency range below $\omega_c$, but sharply drops to a negligibly small value at higher frequencies above $\omega_c$, where  the critical frequency $\omega_c\propto 1/g_{a\gamma^\prime \gamma^\prime}\langle X_{\mu\nu}\rangle$ is determined by the ALP coupling $g_{a\gamma^\prime \gamma^\prime}$ and the strength of the background dark photon gauge field $\langle X_{\mu\nu}\rangle$ (see Fig.~\ref{fig:Benchmarks}). Then the EDGES anomaly can be explained, while avoiding a dangerous CMB distortion, for the model parameters yielding $\omega_{21} < \omega_c < \omega_{\rm CMB}$, which can be achieved when
\bea
\label{condition1}
 4.6\times 10^{-8} \, {\rm GeV}^{-1}\mu{\rm G} \,<\, g_{a\gamma^\prime\gamma^\prime} \langle X_{\mu\nu}\rangle \,<\, 3.7\times 10^{-5} \, {\rm GeV}^{-1}\mu{\rm G}.
\eea

As we need $\langle X_{\mu\nu}\rangle\lesssim 1\,\mu{\rm G}$ to avoid a too large dark radiation energy density, the above condition implies
\bea
\label{condition2}
g_{a\gamma^\prime\gamma^\prime} \,\gtrsim\, 4\times 10^{-8} \, {\rm GeV}^{-1}.\eea
Yet the other ALP couplings $g_{a\gamma\gamma^\prime}$ and $g_{a\gamma\gamma}$ can be small enough to be phenomenologically safe without causing a fine tuning problem. For instance,   $g_{a\gamma\gamma^\prime}$ and $g_{a\gamma\gamma}$
can originate from $g_{a\gamma^\prime\gamma^\prime}$
through a loop-induced kinetic mixing $\varepsilon ={\cal O}(10^{-3}-10^{-2})$ between $\gamma$ and $\gamma^\prime$, which would result in
$g_{a\gamma\gamma}\sim \varepsilon g_{a\gamma\gamma^\prime}\sim \varepsilon^2g_{a\gamma^\prime\gamma^\prime}$.
Alternatively, one may utilize the clockwork mechanism  
\cite{Choi:2014rja, Choi:2015fiu, Kaplan:2015fuy}
 to achieve a hierarchical pattern of ALP couplings, which can generate even an exponentially small  $g_{a\gamma\gamma^\prime}/g_{a\gamma^\prime\gamma^\prime}$ \cite{Higaki:2015jag,Farina:2016tgd,Agrawal:2017cmd}.
Such mechanisms then allow the ALP couplings to satisfy the astrophysical constraints
$g_{a\gamma\gamma^\prime}< 5\times 10^{-10}\,  {\rm GeV}^{-1}$ \cite{Choi:2018mvk} and $g_{a\gamma\gamma}< 5\times 10^{-12}\, {\rm GeV}^{-1}$ \cite{Payez:2014xsa} even when $g_{a\gamma^\prime\gamma^\prime}$ is as large as (\ref{condition2}).

An appealing feature of our scheme is that $P_{\gamma\leftrightarrow a,\gamma^\prime}$ at $\omega_{21}$ 
can be  close to unity over a wide range of model parameters satisfying the observational constraints.
Therefore  our scheme can explain the EDGES anomaly even with a small amount of dark radiations in the 21\,cm frequency range. Another interesting feature  is that the EDGES anomaly can be explained even when the ALP mass $m_a\ll 
m_\gamma(z\simeq 20)\sim 10^{-14}$ eV. Even in such case of ultra-light $m_a$, resonance conversion of DR to 21\,cm photons  can take place
during the period  $20<z<1700$ through an effective DR mass which is determined mostly by $g_{a\gamma^\prime\gamma^\prime} \langle X_{\mu\nu}\rangle$.

 

The organization of this paper is as follows. In the next section, we discuss the resonant conversion between the photon and  dark radiation composed of ALPs and dark photons, which can take place in the early Universe involving a non-zero background dark photon gauge field.    In Sec.~\ref{sec:implication}, we apply this scheme to the EDGES anomaly 
and identify the parameter region which can explain the EDGES signal while satisfying the observational constraints.  We then conclude in Sec.~\ref{sec:con}.
To supplement our discussion, we provide in 
Appendix \ref{app:Overview} a brief summary of the key features of the resonant conversion between the photon and a generic dark radiation in the early Universe; discuss in Appendix \ref{app:generation} an explicit scheme  to generate
a  primordial background dark photon gauge field based on the mechanism of  \cite{Choi:2018dqr}; and finally update in Appendix \ref{app:ALPrevision} the observational constraints on the ALP scenario of \cite{Moroi:2018vci}.



\section{Resonant conversion in ALP-photon-dark photon oscillation scenario}
\label{sec:ResonantConversion}

In this section, we discuss the resonant conversion between $\gamma$ and the dark sector particles composed of ALP and dark photon, which can take place in the early Universe under nonzero background dark photon gauge field.
 In models with an ALP and a dark photon, there can be ALP couplings of the form    \bea
\frac{1}{4}g_{a\gamma \gamma}aF_{\mu\nu}\tilde{F}^{\mu\nu}+
\frac{1}{2}g_{a\gamma \gamma^\prime}aF_{\mu\nu}\tilde{X}^{\mu\nu} +\frac{1}{4}g_{a\gamma^\prime \gamma^\prime}aX_{\mu\nu}\tilde{X}^{\mu\nu},
\label{eq:Lagrangian}
\eea
where $F_{\mu\nu}=\partial_\mu A_\nu-\partial_\nu A_\mu$ is the field strength of the ordinary electromagnetic gauge field $A_\mu$, $\tilde F^{\mu\nu}=\frac{1}{2}\epsilon^{\mu\nu\rho\sigma}F_{\rho\sigma}$ is its dual field strength, and  $X_{\mu\nu}=\partial_\mu X_\nu-\partial_\nu X_\mu$ is the field strength of the $U(1)_X$ dark photon gauge field $X_\mu$. We assume that the dark photon is massless in order to have a long-range cosmic background dark photon gauge field, which is one of the key ingredients of our scheme. This also allows us to avoid $U(1)_X$ breaking sector which may cause a naturalness problem.
Note that for massless $X_\mu$, the kinetic mixing between $A_\mu$ and $X_\mu$ can be rotated away by an appropriate field redefinition of $X_{\mu}$ and $A_\mu$, and the above ALP couplings are defined in such field basis.


In the presence of background dark photon gauge field  \bea
\langle X_{\mu\nu}\rangle =(\vec E_X, \vec B_X),\eea 
the above ALP couplings affect the evolution of the involved particles. We find
the corresponding evolution equation in relativistic limit is given by
\bea
\label{eq:evolution}
\left[i\frac{d}{dt} - \frac{1}{2\omega}\mathcal{M}^2\right] \left(
\begin{array}{c}
A_\parallel \\
X_\parallel \\
a
\end{array}
\right) = 0 \, ,
\eea
where  $A_\parallel$ and $X_\parallel$ denote the polarization states of $A_\mu$ and $X_\mu$ parallel to the background
dark photon gauge field combination
\bea
\vec{\cal B}_X =\vec B_X-\hat k (\hat k\cdot\vec B_X) -\hat k\times \vec E_X,\eea
and the effective mass-square matrix is 
\bea
\mathcal{M}^2=
\left[
\begin{array}{ccc}
m_{\gamma}^2 & 0 & m_{\gamma a}^2 \\
0 & 0 & m_{\gamma^\prime a}^2\\
m_{\gamma a}^2 & m_{\gamma^\prime a}^2 & m_a^2
\end{array}\right]
\label{eq:MassMatrix}
\eea
where
\bea
\label{eq:massmixing}
m_{\gamma a}^2 = g_{a\gamma\gamma^\prime} {\cal B}_X \omega, \quad 
m_{\gamma^\prime a}^2 =  g_{a\gamma^\prime\gamma^\prime}{\cal B}_X\omega \quad \left({\cal B}_X =|\vec {\cal B}_X|\right).
\eea
Here $(\omega, \vec k)$ denote the energy and momentum of the involved particles, $\hat k=\vec k/|\vec k|$,
and we include also the effective photon mass $m_\gamma$  induced by the background thermal plasma in the early Universe.

\begin{figure}[t]
\centering
\includegraphics[width=0.45\textwidth]{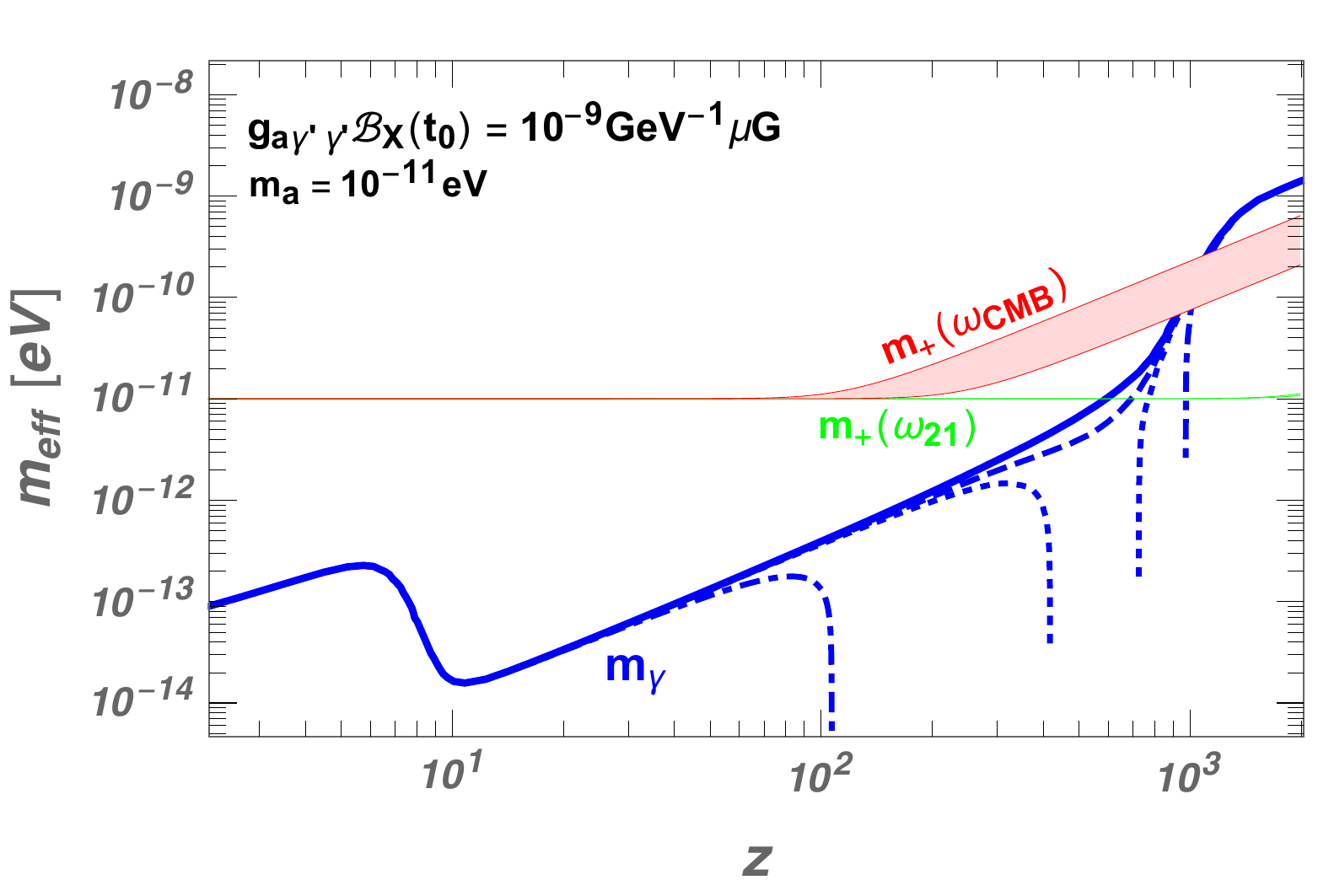} \, \,
\includegraphics[width=0.45\textwidth]{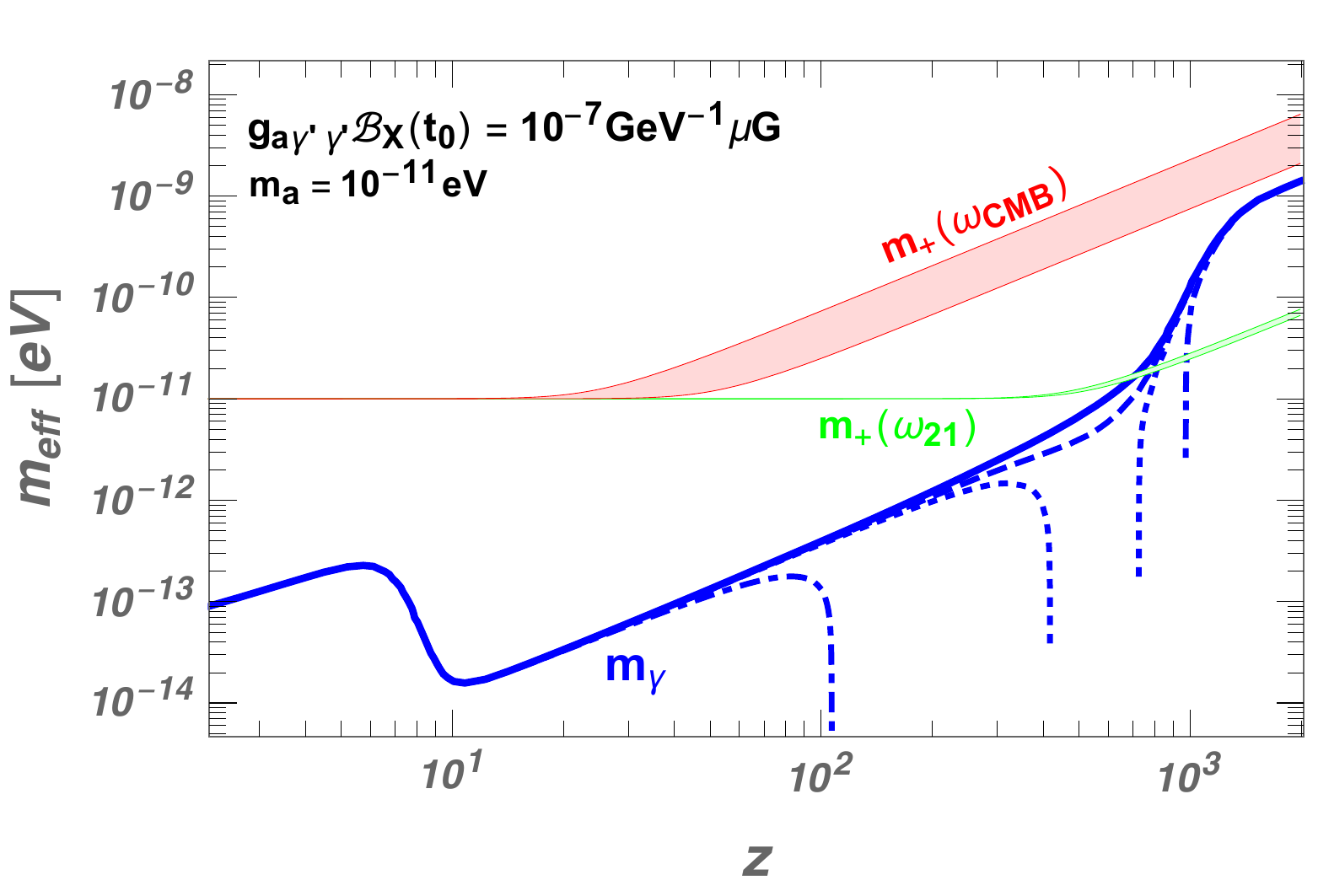} \\
\includegraphics[width=0.45\textwidth]{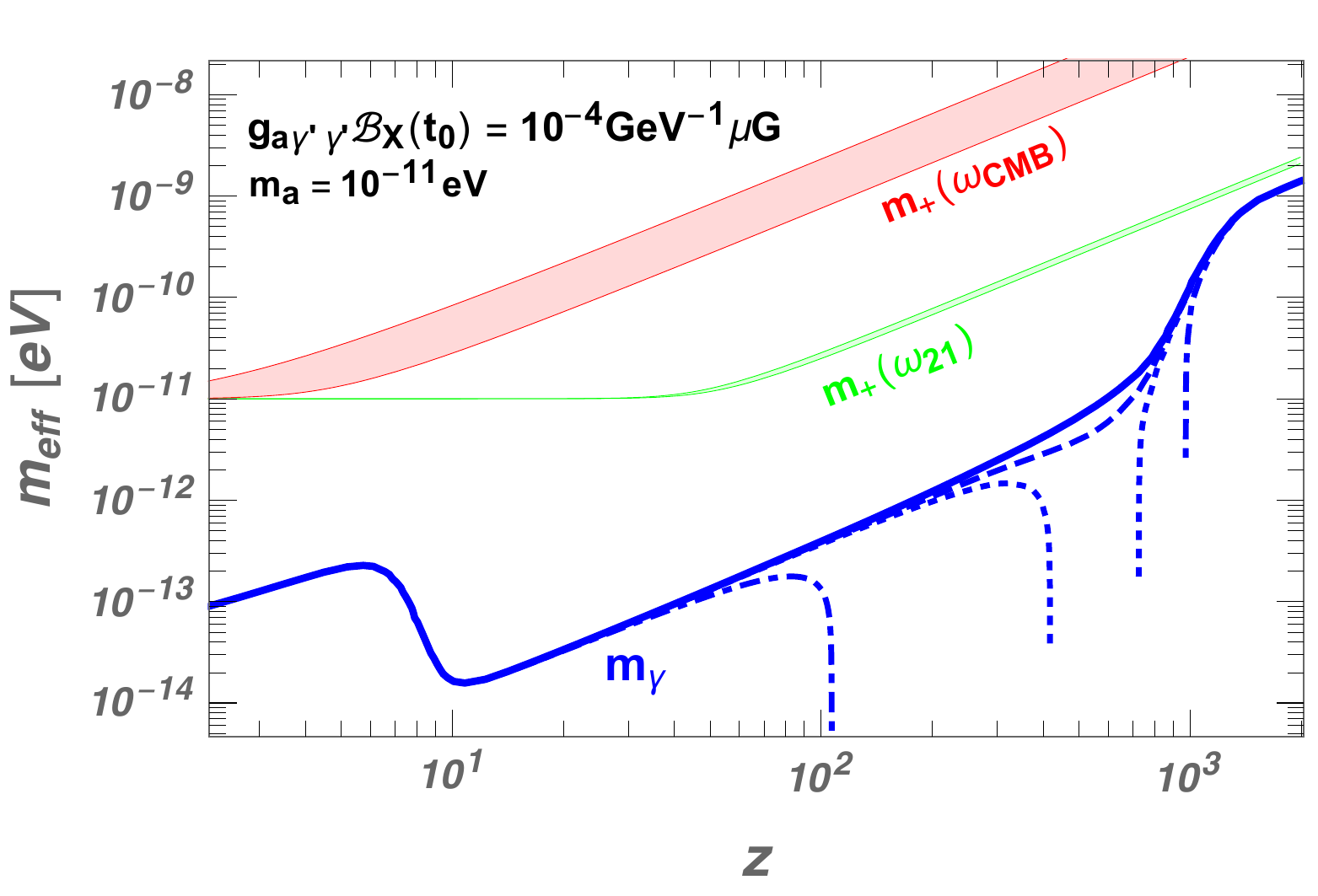} \, \,
\includegraphics[width=0.45\textwidth]{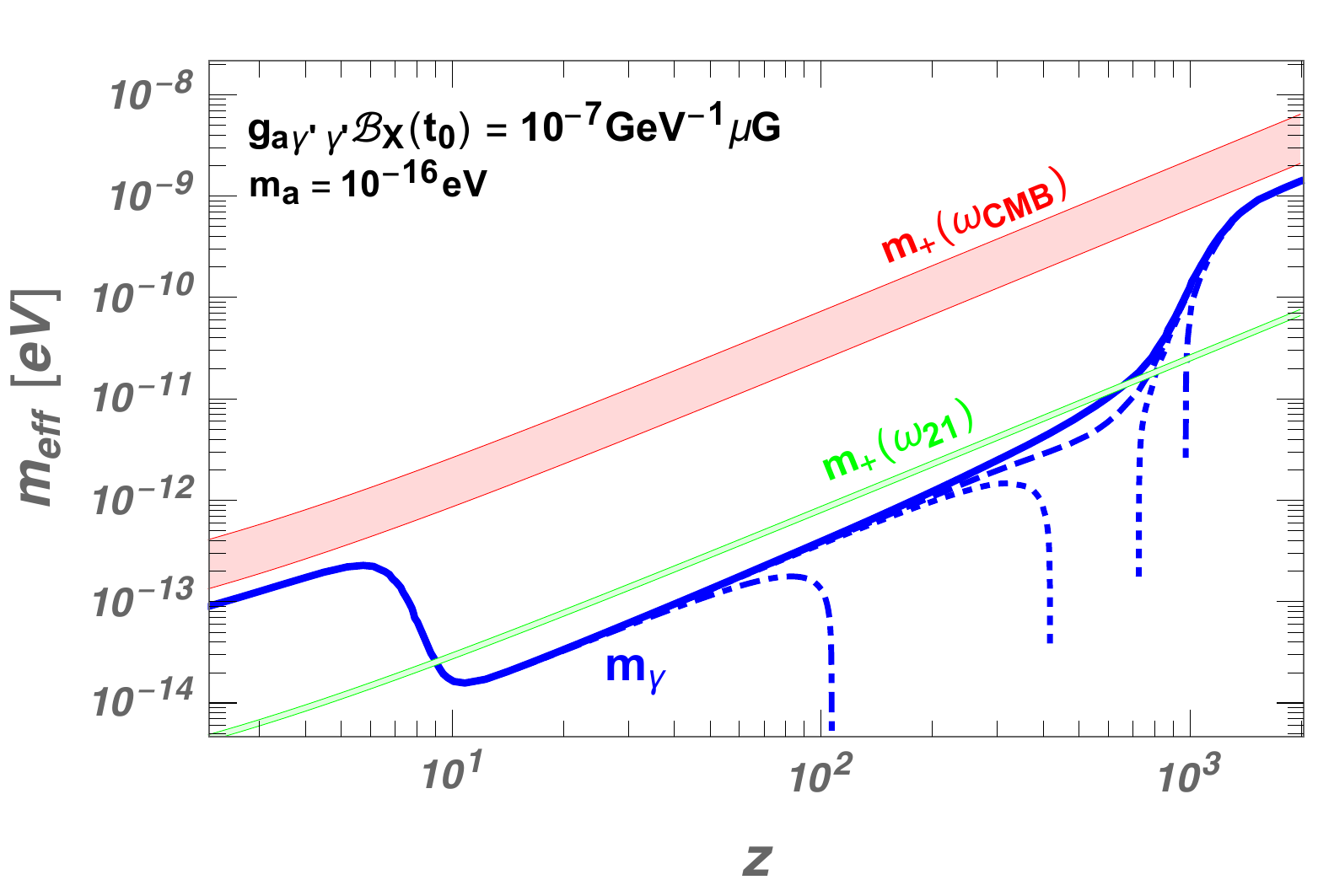}
\caption{Evolution of the effective photon mass (blue) in the early Universe for $\omega=\omega_{21}$(solid), $T$(dashed), $4T$(dotted), $10T$(dash-dotted), where $T$ is the CMB temperature, and also the effective mass of the dark sector mass eigenstate $\phi_+$ with $\omega=\omega_{21}$ (green) and $\omega$ in the COBE-FIRAS frequency range  $[1.2T, \, 11.2T]$ (red). The upper left panel is for
 $(m_a, g_{a\gamma^\prime\gamma^\prime}{\cal B}_X) = (10^{-11}\,{\rm eV}, 10^{-9}\, {\rm GeV}^{-1}\mu{\rm G})$ for which both the CMB and 21\,cm photon experience a resonant conversion, the lower left panel is for
 $(10^{-11}\, {\rm eV},\,  10^{-4}\, {\rm GeV}^{-1}\mu{\rm G})$ for which neither of CMB and 21\,cm photon experiences a resonant conversion, and finally the upper and lower right panels are for $(10^{-11} \,{\rm eV}, 10^{-7}\, {\rm GeV}^{-1}\mu{\rm G})$ and  $(10^{-16}\, {\rm eV}, 10^{-7}\, {\rm GeV}^{-1}\mu{\rm G})$, respectively, for which
 the 21\,cm photon experiences a resonant conversion, while the CMB does not.
}
\label{fig:massSpectrum}
\end{figure}

In order to derive the resonant conversion rate, we need information on the effective photon mass $m_\gamma$.
Here we briefly summarize some features of $m_\gamma$ during the red shift $20<z<1700$. For more details, see Ref.~\cite{Mirizzi:2009iz,Mirizzi:2009nq} and also the blue curves in Fig. \ref{fig:massSpectrum}.
In a circumstance with the hydrogen ionization fraction  $X_e$, the effective photon mass for CMB can be well approximated by \cite{Mirizzi:2009iz,Mirizzi:2009nq}
\bea
m_{\gamma}^2 = \omega_{\rm pl}^2 \times\left[1-7.3\times 10^{-3}\left(\frac{\omega}{\rm eV}\right)^2\left(\frac{1-X_e}{X_e}\right)\right],
\label{eq:EffectivePhotonMass}
\eea
where   $\omega_{\rm pl}^2 = 4\pi\alpha n_e/m_e \simeq 2.53 \times 10^{-28} X_e (1+z)^3 ~{\rm eV}^2$ is the plasma frequency which is 
determined by the electron density $n_e$.
The positive contribution to $m_\gamma^2$, i.e. $\omega_{\rm pl}^2$, originates from the forward scattering off free electrons and the negative contribution \cite{Born:1980} is from the scattering off neutral atoms which can be considered as dielectric medium.
As the negative contribution is proportional to $\omega^2 (1-X_e)$,  it becomes meaningful at high frequency and also when
the neutral hydrogen fraction $1-X_e$ is non-negligible.
Before the recombination with $T\simeq 0.1~{\rm eV}$ and $z\simeq 1100$, the plasma is fully ionized, i.e. $X_e \simeq 1$, so $m_{\gamma}$ is given by $\omega_{\rm pl}$ regardless of $\omega$.
At the recombination,  $X_e$ decreases rapidly to a small value of $\mathcal{O}(10^{-3})$ \cite{Seager:1999bc} and then the negative contribution to $m_\gamma^2$ from neutral atoms can be important.


Here we are interested in the conversion involving $\gamma$, not the conversion just among the dark sector particles.
As we will see, for a successful application of our scheme to the EDGES anomaly,
we need $g_{a\gamma^\prime\gamma^\prime}\gg g_{a\gamma\gamma^\prime}$ and therefore 
$m_{\gamma^\prime a}^2/m_{\gamma a}^2\gg 1$.
It is 
then convenient to rotate away $m_{\gamma^\prime a}^2$ in the mass-square matrix by moving to the instantaneous mass eigenbasis for  dark sector particles. This  can be achieved by the orthogonal 
rotation
\bea
\left(
\begin{array}{c}
\phi_- \\ \phi_+
\end{array}\right) = \left(
\begin{array}{cc}
\cos\theta_D & -\sin\theta_D \\
\sin\theta_D & \cos\theta_D
\end{array}\right)\left(
\begin{array}{c}
X_\parallel \\ a
\end{array}\right),
\eea
where the dark sector mixing angle $\theta_D$ is given by
\bea
\label{eq:tan}
\tan 2\theta_D = \frac{2m_{\gamma^\prime a}^2}{m_a^2}=\frac{2g_{a\gamma^\prime\gamma^\prime}{\cal B}_X\omega}{m_a^2}. 
\eea
In this new basis, the evolution equation is modified as
\bea
\left[i\frac{d}{dt} - \frac{1}{2\omega}\tilde{\mathcal{M}}^2\right] \left(
\begin{array}{c}
A_\parallel \\
\phi_- \\
\phi_+
\end{array}
\right) = 0 \, ,
\eea
where
\bea
\tilde{\mathcal{M}}^2 = 
\left(
\begin{array}{ccc}
m_{\gamma}^2 & -m_{\gamma a}^2\sin\theta_D & m_{\gamma a}^2\cos\theta_D \\
-m_{\gamma a}^2\sin\theta_D & m_-^2 &2\omega\dot\theta_D\\
m_{\gamma a}^2\cos\theta_D & 2\omega\dot\theta_D & m_+^2
\end{array}\right) \eea
for the instantaneous dark sector mass eigenvalues 
\bea
m_\pm^2 = \frac{m_a^2}{2}\pm \frac{\sqrt{m_a^4+4m_{\gamma^\prime a}^4}}{2} \, . \eea 
In our case, \bea
|\omega\dot\theta_D|\,\ll\, |m_+^2-m_-^2|, \quad  m_{\gamma a}^2 < m_\gamma^2\eea over the frequency range and cosmic period relevant for us. In such case, we can safely ignore the components $(\tilde {\cal M}^2)_{23}=(\tilde {\cal M}^2)_{32}=2\omega\dot\theta_D$ as they do not significantly affect the evolution of dark sector particles.
Then our problem is reduced to a resonance conversion between $\gamma$ and $\phi_\pm$, which is induced by $m_{\gamma a}^2$
 when the resonance condition  $m_\gamma^2= m_\pm^2$ is fulfilled.
In Appendix~\ref{app:Overview}, we briefly summarize the key features of the resonant conversion between $\gamma$ and a generic dark sector particle $\phi$ which can have time-dependent mass $m_\phi$ in the early Universe.

\subsection{Small dark sector mixing}
\label{subsec:smalldarkmixing}

In the limit $g_{a\gamma^\prime\gamma^\prime}{\cal B}_X \omega \ll m_a^2$, 
the dark sector mixing angle $\theta_D \ll 1$. In this limit, the propagation eigenstates are given by 
\bea \phi_+\simeq a,  \quad  \phi_-\simeq X_\parallel, \eea 
with the mass eigenvalues
\bea    m_+^2\simeq m_a^2,  \quad m_-^2\simeq -\theta_D^2 m_a^2.\eea 
 Then  resonant conversion can take place between $\gamma$ and $\phi_+$ when $m_\gamma^2(z)=m_+^2$ in the early Universe\footnote{For $w>3.8 \, T$, due to the scattering off by neutral atoms, 
$m_\gamma^2$ can become negative for  a while within the period $20 < z < 1100$ \cite{Mirizzi:2009iz,Mirizzi:2009nq}. Then there can be a resonant conversion between $\gamma$ (with $\omega > 3.8 \, T$) and $\phi_-\simeq  X_\parallel$  when  the resonance condition $m_\gamma^2=m_-^2\simeq -\theta_D^2 m_a^2$ is
fulfilled. However, in such case $m_{\gamma}^2$ is sharply varying  at the resonance point, and as a consequence the conversion probability is suppressed by the small factor $m_-^2/m_a^2\sim \theta_D^2$.}.
Since $m_+^2\simeq m_a^2$ is approximately a constant,  this is essentially same as the well known $\gamma$-$a$  conversion induced by the conventional ALP coupling $g_{a\gamma\gamma} aF\tilde F$ in background magnetic field $B$, 
but with $g_{a\gamma\gamma}B$ replaced by $g_{a\gamma\gamma^\prime}{\cal B}_X$. 
Obviously, in this case the resonance condition $m_\gamma^2(z)=m_+^2\simeq m_a^2$ can be fulfilled for $\omega=\omega_{21}$ and $20<z<1700$ only for 
\bea
m_a={\cal O}(10^{-14}-10^{-9})\,\, {\rm eV}.
\label{eq:ALPmassSmallMixing}
\eea

For subsequent discussions, it is convenient to define  
\bea
\label{eq:omegaL}
\omega_L(t_0) =\frac{1}{(1+z_{\rm res})^3}\frac{m_a^2}{g_{a\gamma^\prime\gamma^\prime}{\cal B}_{X}(t_0)}, 
\eea
where $t_0$ denotes the present Universe and $z_{\rm res}$ is the red-shift at the time $t=t_{\rm res}$ when the resonance condition $m_\gamma^2 =m_+^2\simeq m_a^2$ is met.
Note that for  $m_a < 1.6\times 10^{-14}$ eV, such  resonance condition can not be fulfilled, so $\omega_L$ for $m_a < 1.6\times 10^{-14}$ eV is fixed to the value at $m_a=1.6\times 10^{-14}~{\rm eV}$ and $z_{\rm res} = 11$.
Then the resonant conversion between $\gamma$ and $\phi_+\simeq a$ takes place in the small dark sector mixing regime for the ALP mass range \eqref{eq:ALPmassSmallMixing} and the frequency
\bea
\omega \,\ll\, \omega_L,\eea
where  we assume that the background dark photon gauge field is generated before $t_{\rm res}$ and subsequently red-shifted 
 as ${\cal B}_X (t_0) = {\cal B}_X(t)/(1+z)^2$.
 The corresponding resonant conversion probability can be obtained from Eqs.~\eqref{prob1} and \eqref{prob2} by inserting $m_{\rm mix}^2 = g_{a\gamma\gamma^\prime}{\cal B}_X\omega$ and $m_\phi^2=m_a^2$, which results in
\bea
\label{eq:prob_small_mixing}
P_{\gamma\leftrightarrow  \phi_+\simeq a}(\omega\ll \omega_L)\, \simeq\,  1-p_{\rm res}\,\simeq\,
 1-\left.\exp\left(-r\frac{\pi g_{a\gamma\gamma^\prime}^2 {\cal B}_X^2\omega}{m_a^2}\right)\right|_{t=t_{\rm res}},
\eea
where 
\bea
r^{-1}={|d\ln (m_{\gamma}^2/m_{+}^2)/{dt}|_{t=t_{\rm res}}}={\cal O}\left(1-10\right)\times H(t_{\rm res})\eea
for the Hubble expansion rate $H(t)$.
In the parameter region giving $r
g_{a\gamma\gamma^\prime}^2{\cal B}_X^2 \omega/m_a^2 \gtrsim 1$,   
the conversion probability is close to unity and nearly independent of the photon frequency $\omega$. On the other hand, in the other limit with 
$rg_{a\gamma\gamma^\prime}^2{\cal B}_X^2\omega/m_a^2\ll 1$,  the conversion probability is small and proportional to the photon frequency as
\bea
\label{eq:smallprob_small_mixing}
P_{\gamma\leftrightarrow  \phi_+\simeq a} (\omega\ll \omega_L)\, \simeq\, r\frac{\pi g_{a\gamma\gamma^\prime}^2 {\cal B}_X^2\omega}{m_a^2}.\eea
As the dark photon does not participate in resonant conversion in this case,  
the photon density spectrum is reshaped mainly by the $\gamma$-$a$ conversion as
\bea
\frac{dn_\gamma}{d\omega} \quad \rightarrow \quad \frac{dn_\gamma}{d\omega} \times \left(P_{\gamma\rightarrow\gamma}\right)_{\omega\ll\omega_L}+ \frac{dn_a}{d\omega} \times \left(P_{\gamma\rightarrow \phi_+\simeq a}\right)_{\omega\ll\omega_L} \, ,
\eea
where
$P_{\gamma\rightarrow\gamma}\,\simeq\, 1-P_{\gamma\rightarrow \phi_+\simeq a}$ is the photon survival probability.




\subsection{Large dark sector mixing}
\label{subsec:largedarkmixing}

\begin{figure}[t]
\centering
\includegraphics[width=0.6\textwidth]{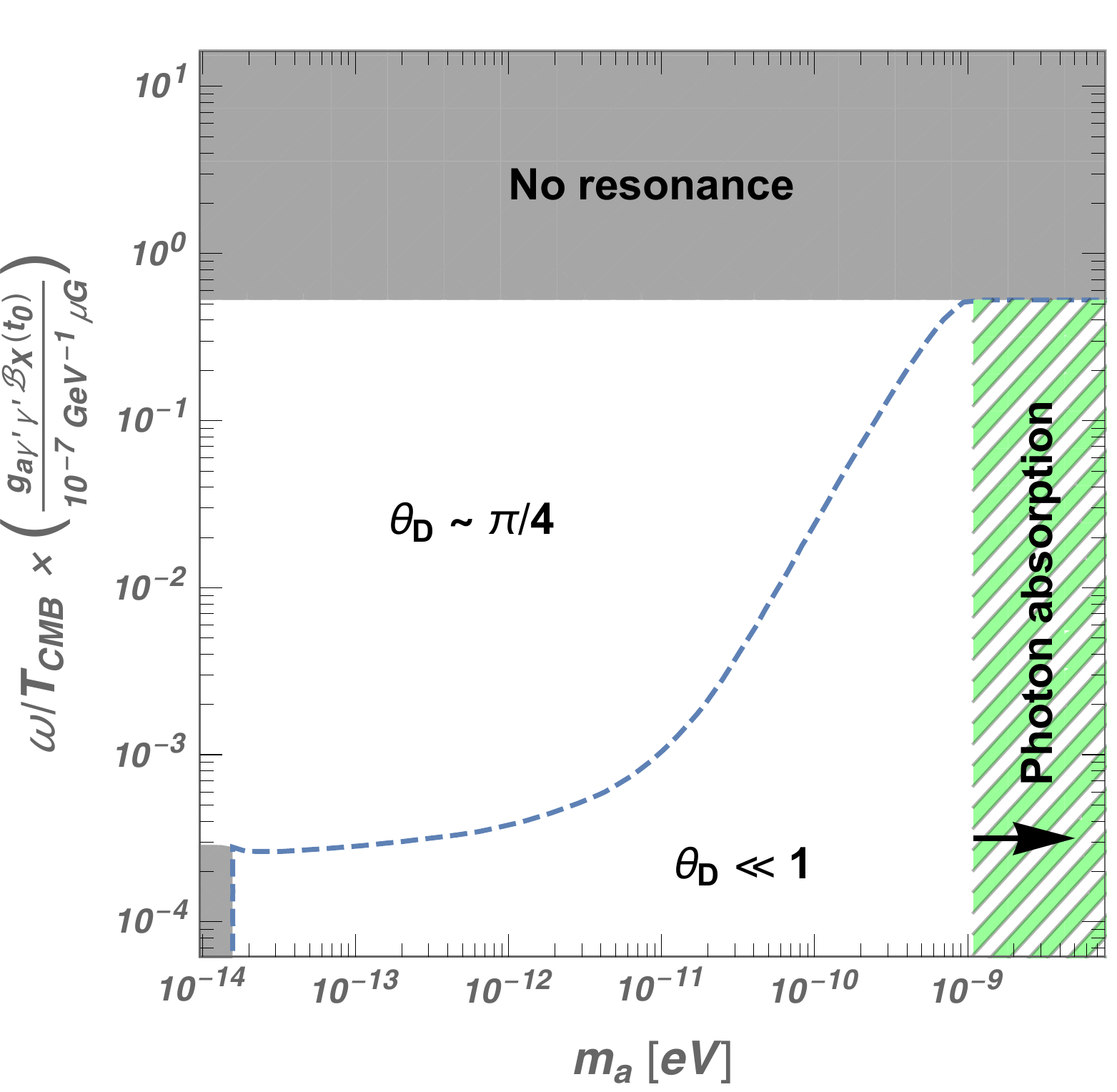}
\caption{Parameter regions for small dark sector mixing, large dark sector mixing, and no resonance. The dotted curve corresponds to $\tan2\theta_D=1$, and the gray area does not allow the resonance condition $m_\gamma^2=m_\pm^2$ to be fulfilled.}
\label{fig:FrequencySpectrum}
\end{figure}

The most interesting feature of our scheme appears
in the  large dark sector mixing regime with $\tan 2\theta_D = 2 g_{a\gamma^\prime\gamma^\prime}{\cal B}_X \omega/m_a^2\gg 1$. In such case, the propagation eigenstates are given by the nearly maximal mixtures of ALP and dark photon,
\bea
\phi_{\pm}\simeq  \frac{X_\parallel\pm a}{\sqrt{2}}\eea with
 the mass eigenvalues 
\bea
m_\pm^2 \simeq \pm m_{\gamma^\prime a}^2 =\pm g_{a\gamma^\prime\gamma^\prime} {\cal B}_X\omega \, .
\eea 
In this case also, the primary resonance conversion takes place between $\gamma$ and $\phi_+$ when $m_\gamma^2=m_+^2$. However there is a 
key difference from the small dark sector mixing case.
The mass eigenvalue $m_+^2$ in the large mixing case is red-shifted as $(1+z)^3$ in the expanding Universe, 
while it is approximately constant in the small mixing case.

More specifically, $m_\pm^2$ in the limit $\tan 2\theta_D\gg 1$ is red-shifted like   $\omega_{\rm pl}^2 \propto n_e \propto (1+z)^3$ when the hydrogen ionization fraction $X_e$ is constant, e.g. before the recombination and after the re-ionization.
(See for instance the cosmic evolution of $m_+$ and $m_\gamma$ in Fig.~\ref{fig:massSpectrum}.)
By virtue of this coincidence, if $g_{a\gamma^\prime \gamma^\prime} {\cal B}_{X}\omega$ is large enough, $m_+^2 > \omega_{\rm pl}^2$ over the whole evolution history, so the resonance condition $m_\gamma^2=m_+^2$ can never be fulfilled as in
the case of $m_+(\omega_{\rm CMB})$ in the upper right panel and lower two panels of Fig.~\ref{fig:massSpectrum}. 
One easily finds that this happens for 
\bea
\omega(t_0) > \omega_c(t_0)=1.4 \times 10^{-4}~{\rm eV} \left(\frac{10^{-7}~{\rm GeV}^{-1}\mu{\rm G}}{g_{\gamma^\prime\gamma^\prime}{\cal B}_{X}(t_0)}\right)\,, 
\label{eq:DarkLargeMixingRange}
\eea
which corresponds to the upper gray region in Fig.~\ref{fig:FrequencySpectrum}.
In Fig.~\ref{fig:FrequencySpectrum}, we depict also the curve for $\tan 2\theta_D =1$ which splits 
the parameter space 
into the two regions,  the large dark sector mixing region (upper) and the small dark sector mixing region (lower).

If the critical frequency $\omega_c$ is lower than the CMB frequencies $\omega=[1.2T, \,11.2T]$ probed by the COBE-FIRAS \cite{Fixsen:1996nj}, i.e. \bea
\omega_c(t_0) < 2.84\times 10^{-4} ~ {\rm eV}\eea
or equivalently
\bea
\label{eq:no-resonance-condition}
g_{a\gamma^\prime \gamma^\prime} {\cal B}_{X}(t_0) > 4.6\times 10^{-8}~ {\rm GeV}^{-1} \mu {\rm G} \, , \label{eq:KeyResult}
\eea
 there could be no resonant conversion for the CMB in the COBE-FIRAS frequencies
 as in the upper right panel and lower two panels of Fig.~\ref{fig:massSpectrum},
which is one of the most interesting features of our scheme.  
We note that once the above condition is satisfied, which assures that $\phi_+$ does not experience a resonance conversion to
$\gamma$ in the COBE-FIRAS frequency range,  $\phi_-$ also can not have a resonance conversion to $\gamma$ in the same frequency range.
Even though $m_\gamma^2$ for $\omega> 3.8 \,T$ becomes negative during certain period as in the case of dotted and dash-dotted blue curves in Fig.~\ref{fig:massSpectrum}, 
its absolute value is not large enough 
to satisfy $m_\gamma^2 =m_-^2 \simeq -g_{a\gamma^\prime\gamma^\prime} {\cal B}_X\omega$ 
for the COBE-FIRAS frequencies and $g_{a\gamma^\prime\gamma^\prime} {\cal B}_X$ satisfying (\ref{eq:KeyResult}).
Yet, there can be {\it non-resonant} conversion between CMB and $\phi_{\pm}$. As there is no resonant conversion,
we can set $p_{\rm res}=0$ in (\ref{prob1}), and find 
the corresponding non-resonant conversion probability 
\bea
\label{eq:non-resonant-prob}
P_{\gamma\leftrightarrow a,\gamma^\prime}(\omega>\omega_c)\,\simeq\, \frac{1}{2}  \frac{g^2_{a\gamma\gamma^\prime}}{g^2_{a\gamma^\prime\gamma^\prime}}
\eea 
for $g_{a\gamma\gamma^\prime}/g_{a\gamma^\prime\gamma^\prime}\ll 1$.



Due to the rapid reduction of the hydrogen ionization fraction $X_e$, the effective photon mass
$m_\gamma^2$  is more rapidly red-shifted than $m_+^2\simeq g_{a\gamma^\prime\gamma^\prime}{\cal B}_{X}\omega\propto (1+z)^3$
right after the recombination (see 
Fig.\,\ref{fig:massSpectrum}). As a consequence,
there exists an intermediate frequency range over which the resonance condition $m_\gamma^2=m_+^2$ can be fulfilled in the large dark sector mixing regime  
as in
the case of $m_+(\omega_{21})$ in the upper and lower right panels  of Fig.~\ref{fig:massSpectrum}. 
Such frequency range is given by
\bea
\omega_L(t_0)\, < \,\omega (t_0)\, < \, \omega_c(t_0),  
\label{eq:LargeMixingResonance}
\eea
where $\omega_c$  is given in (\ref{eq:DarkLargeMixingRange}) and
$\omega_L$  
 is  given  in (\ref{eq:omegaL}).
In Fig.~\ref{fig:FrequencySpectrum}, the parameter region above the dotted curve but below the upper gray area corresponds to this intermediate frequency range.
In fact, in this case there can be a resonant conversion between $\gamma$ and $\phi_+$ 
during the period $20<z<1700$ even when $m_a\ll m_\gamma(z=20)\sim 10^{-14}$ eV. This is because the resonance conversion occurs through the effective mass $m^2_+\simeq g_{a\gamma^\prime\gamma^\prime}{\cal B}_X\omega$  satisfying $m_+^2=m_\gamma^2$  even when  $m_a\ll 10^{-14}$ eV.
In such case, as shown in the lower right panel of Figure.~\ref{fig:massSpectrum}, there is an additional resonant conversion at lower redshift $z < 10$.
However  this later conversion is less significant than the earlier one occurring at $z\sim 10^3$ because it is less adiabatic.

\begin{figure}[t]
\centering
\includegraphics[width=0.65\textwidth]{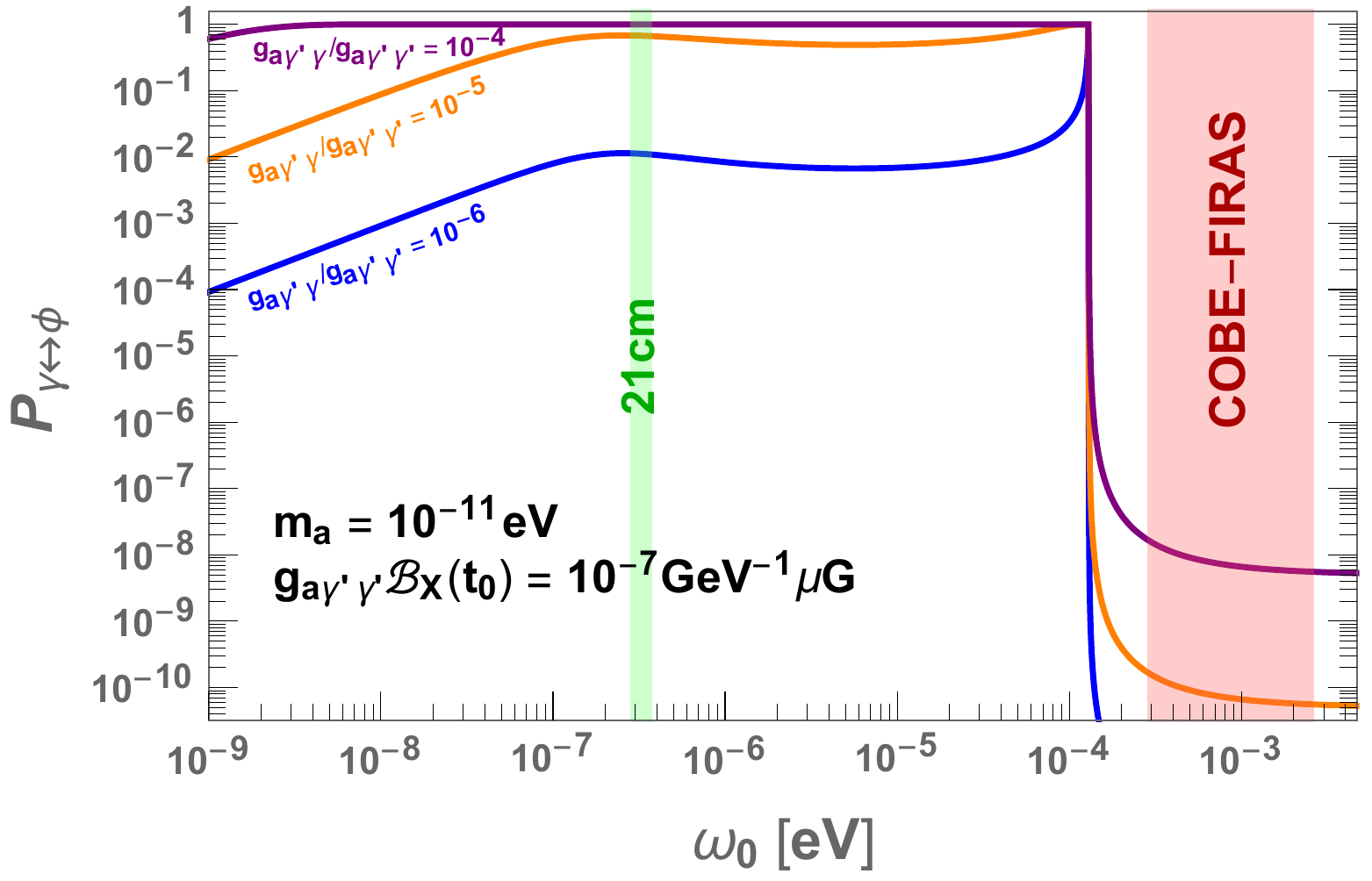}
\caption{Spectral dependence of the resonant conversion probability $P_{\gamma\leftrightarrow \phi_+}$ for the  parameter choice of $m_a = 10^{-11}{\rm eV}$, $g_{a\gamma^\prime\gamma^\prime}{\cal B}_X(t_0) = 10^{-7}{\rm GeV}^{-1}\mu{\rm G}$, and $g_{a\gamma\gamma^\prime}/g_{a\gamma^\prime\gamma^\prime} = 10^{-6}$ (blue), $10^{-5}$ (orange), and $10^{-4}$ (purple). The COBE-FIRAS frequency range corresponds to the red-colored region where the resonance condition can not be fulfilled, and therefore the conversion occurs through tiny non-resonant process.  The conversion probability
$P_{\gamma\leftrightarrow \phi_+}\propto \omega$ in
the low frequency regime with $\omega < \omega_L$, nearly flat over $\omega_L<\omega <\omega_c$, and then sharply drop to a small
non-resonant conversion probability $P_{\gamma\leftrightarrow \phi_+}\simeq g_{a\gamma\gamma^\prime}^2/2g_{a\gamma^\prime\gamma^\prime}^2$ at $\omega>\omega_c$.}
\label{fig:Benchmarks}
\end{figure}

Applying Eqs.~\eqref{prob1} and \eqref{prob2} for the mass parameters \bea
m_\phi^2  \simeq  m_{\gamma^\prime a}^2=g_{a\gamma^\prime\gamma^\prime} {\cal B}_X\omega, \quad  m_{\rm mix}^2  =  -\frac{m_{\gamma a}^2}{\sqrt{2}}= -\frac{g_{a\gamma\gamma^\prime} {\cal B}_X\omega}{\sqrt{2}},\eea 
while assuming $g_{a\gamma\gamma^\prime} \ll g_{a\gamma^\prime\gamma^\prime}$,
we find that the conversion probability in the intermediate frequency range is given by
\bea
P_{\gamma\leftrightarrow a,\gamma^\prime}(\omega_L<\omega <\omega_c)\,\simeq\, 1- p_{\rm res}
\,\simeq\, 1-\left.\exp\left(-r\frac{\pi g_{a\gamma\gamma^\prime}^2 {\cal B}_X}{2 g_{a\gamma^\prime\gamma^\prime}}
\right)\right|_{t=t_{\rm res}}, 
\label{eq:ConvRateLargeDarkMixing}
\eea
where
\bea
r^{-1}={|d\ln (m_{\gamma}^2/m_{+}^2)/{dt}|_{t=t_{\rm res}}}={\cal O}\left(1-10\right)\times H(t_{\rm res}).\eea
Note that the resonance conversion in this regime  typically occurs right after the recombination when $X_e$ rapidly decreases
as in the case of
$m_+(\omega_{21})$ in the upper and lower right panels of Fig.~\ref{fig:massSpectrum}.
 As a consequence, the frequency-dependence of the conversion probability is weakened and becomes approximately $\omega$-independent as in (\ref{eq:ConvRateLargeDarkMixing}), which is
another interesting feature of our scheme. We note also that the conversion probability is quite sensitive to
the ALP coupling ratio $g_{a\gamma\gamma^\prime}/g_{a\gamma^\prime\gamma^\prime}$. 
The above resonance conversion between $\gamma$ and $\phi_+$ in the large dark sector mixing regime   results in the modification of the photon density spectrum 
as
\bea
\frac{d n_\gamma}{d\omega}  \quad \rightarrow \quad
 \frac{dn_\gamma}{d\omega}\times \left(P_{\gamma\rightarrow \gamma}\right)_{\omega>\omega_L}  +
 \frac{1}{2}\left(\frac{dn_a}{d\omega}+\frac{dn_{\gamma^\prime}}{d\omega}\right) \times \left(P_{\gamma\leftrightarrow a,\gamma^\prime}\right)_{\omega>\omega_L} \, ,
\eea
where the factor $1/2$ originates from the large mixing between dark photon  and ALP.
Note that although $\gamma^\prime$ is exactly massless in the vacuum, it has a large mixture with the (approximate) mass eigenstate $\phi_+$ when $g_{a\gamma^\prime\gamma^\prime}{\cal B}_X\omega \gtrsim m_a^2$, so actively
participates in the resonant conversion
to $\gamma$ in the early Universe.

In Fig.~\ref{fig:Benchmarks}, we depict the spectral dependence of the conversion probability  $P_{\gamma\leftrightarrow \phi_+}$ for $m_a=10^{-11}\,{\rm eV}$, $g_{a\gamma^\prime\gamma^\prime}{\cal B}_X = 10^{-7}~{\rm GeV}^{-1}\mu{\rm G}$, and three different values of
$g_{a\gamma\gamma^\prime}/g_{a\gamma^\prime\gamma^\prime} = 10^{-4}$, $10^{-5}$ and $10^{-6}$.
As was anticipated from (\ref{eq:prob_small_mixing}) and (\ref{eq:smallprob_small_mixing}), in the low frequency regime with $\omega \ll \omega_L$, which corresponds to the small dark sector mixing regime, the conversion probability has a nearly flat spectral  dependence when it is close to unity, which is the case for $g_{a\gamma\gamma^\prime}/g_{a\gamma^\prime\gamma^\prime} = 10^{-4}$, but
grows as $P_{\gamma\leftrightarrow \phi_+} \propto \omega$ when it is small, which is the case for $g_{a\gamma\gamma^\prime}/g_{a\gamma^\prime\gamma^\prime} = 10^{-5}, 10^{-6}$.  As (\ref{eq:ConvRateLargeDarkMixing}) indicates, the conversion probability
is approximately flat over the intermediate frequency range
$\omega_L<\omega<\omega_c$ regardless of the value of $g_{a\gamma\gamma^\prime}/g_{a\gamma^\prime\gamma^\prime}$,
and finally sharply drops to a small non-resonant conversion probability $P_{\gamma\leftrightarrow \phi_+}\simeq 
g_{a\gamma\gamma^\prime}^2/2g_{a\gamma^\prime\gamma^\prime}^2$ 
 at $\omega>\omega_c$. For  the chosen values of $m_a$ and $g_{a\gamma^\prime\gamma^\prime}{\cal B}_X$, 
we have $\omega_{\rm CMB}>\omega_c$ (red-colored) and $\omega_L< \omega_{21}< \omega_c$ (green-colored).
Fig.~\ref{fig:Benchmarks} shows that our scheme can give a sizable conversion of dark radiations, either ALPs or dark photons, to 21\,cm photons, even with a probability close to unity, while avoiding dangerous CMB distortions.


\section{Implication for the EDGES 21\,cm signal}
\label{sec:implication}

The ALP-photon-dark photon oscillation discussed in the previous section provides an appealing mechanism to 
 explain the recent tentative observation by the EDGES experiment  of an
 anomalously strong absorption signal of 21\,cm photons \cite{Bowman:2018yin}.
In this section, we examine 
this possibility in more detail.


To explain the EDGES anomaly by a resonant conversion of 
  $\phi$ into 21\,cm photons \cite{Pospelov:2018kdh,Moroi:2018vci}, we need 
a conversion probability  \bea
P_{\phi\rightarrow \gamma} \,\sim\, \frac{10^{-9}}{f_\phi^{\rm 21cm}\Delta N_{\rm eff}^{\phi}},\eea
 where $\Delta N_{\rm eff}^{\phi}$ denotes the energy density of $\phi$ parametrized by the effective number of neutrino species and $f_\phi^{\rm 21cm}$ is the energy fraction in the 21\,cm frequency range.
 Since the energy density of total dark radiation is bounded as
  $\Delta N_{\rm eff} \lsim 0.3$ \cite{Ade:2015xua}, the conversion probability has to be at least of the order of $10^{-8}$ to explain the EDGES anomaly.
For the scenarios proposed \cite{Pospelov:2018kdh,Moroi:2018vci}, the parameter region giving  $P_{\phi\rightarrow \gamma} \gtrsim 10^{-8}$ is
severely limited by a variety of phenomenological constraints. As a consequence, either only a tiny parameter region remains to be viable,  
or the viable parameter region may suffer from a fine-tuning problem.
For instance,  for the dark photon scenario of \cite{Pospelov:2018kdh}, generating the tiny dark photon mass
$m_{\gamma^\prime}={\cal O}(10^{-14}-10^{-9})$ eV and also small kinetic mixing $\epsilon < 10^{-5}$ in the UV completion of the model
may cause a naturalness problem. 
As for the ALP scenario of \cite{Moroi:2018vci},  we combine in Appendix~\ref{app:ALPrevision} the CMB distortion constraint  with the additional constraints from the absence of $\gamma$-ray burst associated with SN1987A \cite{Payez:2014xsa} and the upper bound 
on the primordial background magnetic field  to avoid an overheating of baryons which would wash away the EDGES signal \footnote{Since we regard the EDGES result as a signal of 21\,cm absorption, we take this bound on the primordial magnetic field as a real constraint.} \cite{Minoda:2018gxj}. 
We then find that the parameter region giving $P_{\phi\rightarrow \gamma}\gtrsim  10^{-7}$ for the ALP mass  $m_a=10^{-14}-10^{-9}$ eV is fully excluded by these constraints.  
If the primordial background magnetic field nearly saturates its upper bound 
$B_0\lesssim 0.1$ nG, a tiny parameter region can give $P_{\phi\rightarrow \gamma} \gtrsim 10^{-8}$, while satisfying the  observational constraints.  
This means that the ALP scheme of \cite{Moroi:2018vci} can explain the EDGES anomaly {\it only} when both the ALP dark radiation and 
the primordial background magnetic field nearly saturate their upper bounds, i.e.
 $f_\phi^{\rm 21cm}\Delta N_{\rm eff}^{\phi}\sim 0.1$ and  $B_0\sim 0.1$ nG.    
On the other hand, our scenario can give  a large conversion probability even close to  unity, while satisfying the observational constraints and also without causing a fine-tuning problem.
As a result, in our scheme 
even a small amount of dark radiation at the 21\,cm frequency range, e.g. $f_\phi^{\rm 21cm}\Delta N_{\rm eff}^{\phi}\sim 10^{-9}$, can give an enough boost to heat up the 21\,cm photons, so can explain the EDGES anomaly.


\begin{figure}[t]
\centering
\includegraphics[width=0.6\textwidth]{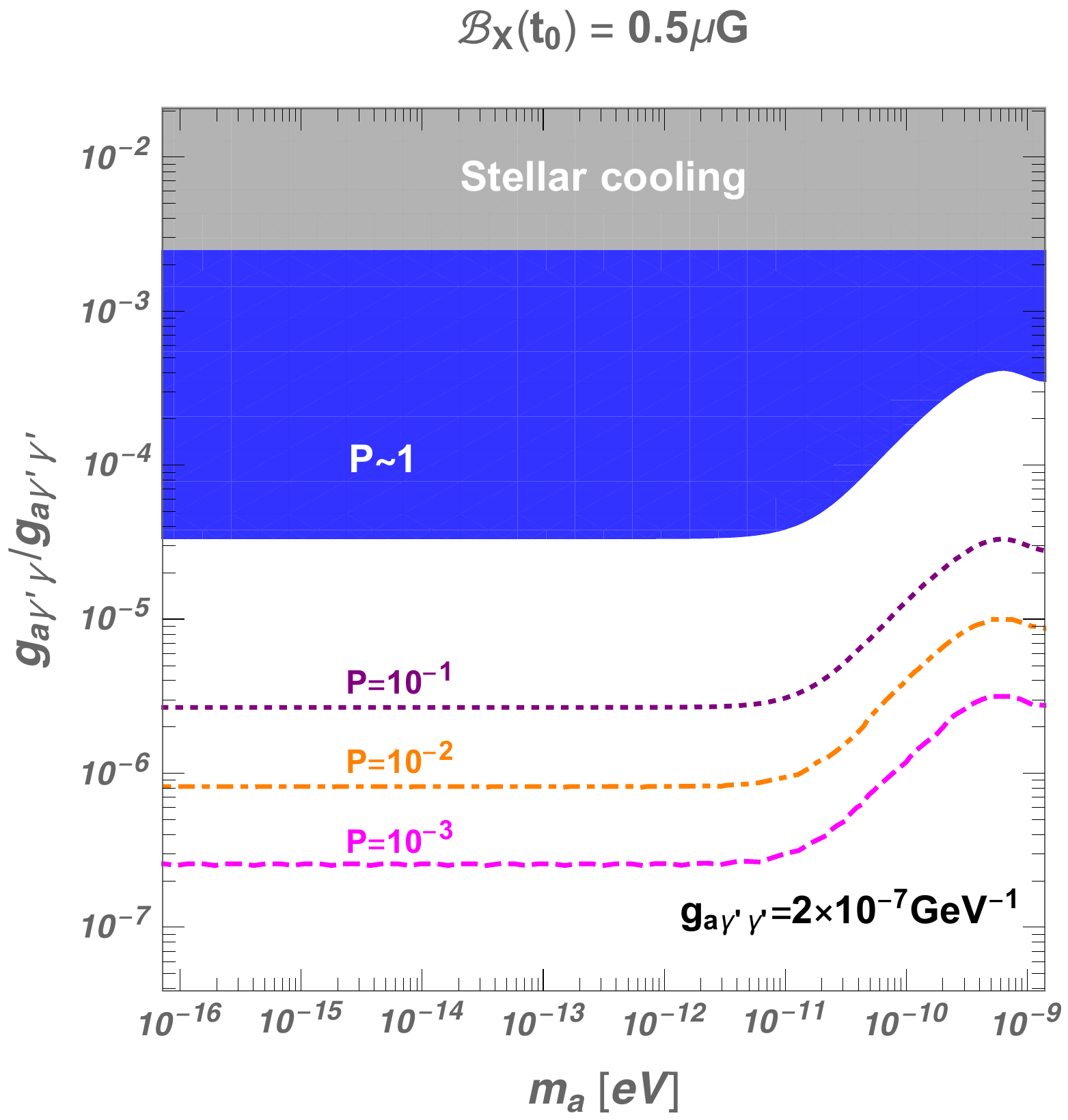}
\caption{Contour of the resonant conversion probability $P_{\gamma\leftrightarrow \phi_+} \simeq 1$ (blue region), $10^{-1}$ (purple), $10^{-2}$ (orange), $10^{-3}$ (magenta) for 21\,cm photons. Here we  take  $g_{a\gamma^\prime\gamma^\prime} = 2\times 10^{-7}\,{\rm GeV}^{-1}$ and ${\cal B}_X(t_0) = 0.5 \,\mu{\rm G}$. The gray region is excluded by the stellar cooling bound $g_{a\gamma\gamma^\prime} < 5\times 10^{-10}\, {\rm GeV}^{-1}$.}
\label{fig:Pres21cm}
\end{figure}

Let us now identify the model parameter region of our scheme which can explain the EDGES anomaly while satisfying the observational constraints.
To have a resonant conversion of dark radiations to 21\,cm photons at $20 < z< 1700$, we first need\bea
  m_a < 10^{-9}\, {\rm eV}.\eea
  Note that in our scheme such resonance conversion can occur even when $m_a\ll m_\gamma(z\simeq 20)\sim 10^{-14}$ eV as long as the effective DR mass  fulfils  the resonance condition 
  $m_+^2 \simeq g_{a\gamma^\prime\gamma^\prime}{\cal B}_X \omega=m_\gamma^2$ for $\omega=\omega_{21}$ and $20<z<1700$.
As was noticed in the previous section, we can  avoid a resonant conversion at the CMB frequencies,
while having a large conversion at the 21\,cm frequency, by arranging the model parameters to have $\omega_{21} < \omega_c <\omega_{\rm CMB}$,
where $\omega_c$ is the critical frequency given by (\ref{eq:DarkLargeMixingRange}).
This can be achieved for
\bea
\label{condition1}
 4.6\times 10^{-8} \, {\rm GeV}^{-1}\mu{\rm G} \,<\, g_{a\gamma^\prime\gamma^\prime} {\cal B}_X \,<\, 3.7\times 10^{-5} \, {\rm GeV}^{-1}\mu{\rm G},
\eea
where we used $\omega_{\rm CMB}\simeq 1.2 \,T_{\rm CMB}$ which corresponds to  the lowest CMB frequency probed by the COBE-FIRAS. As the background dark photon gauge field $\langle X_{\mu\nu}\rangle \propto (1+z)^2$, its energy density is a part of the total dark radiation energy density which is bounded as
$\Delta N_{\rm eff} <  0.3$ \cite{Ade:2015xua}. This requires that ${\cal B}_X \lesssim 1~\mu{\rm G}$,
and then the above condition implies that we need
\bea
\label{bound_2}
g_{a\gamma^\prime\gamma^\prime} \,\gtrsim \,  4\times 10^{-8} \, {\rm GeV}^{-1}.\eea
We note that practically there is no observational constraint on $g_{a\gamma^\prime\gamma^\prime}$, so in principle 
$g_{a\gamma^\prime\gamma^\prime}$ can be  significantly bigger than the above lower bound.
Absence of a resonant conversion at CMB frequencies does not guarantee that the scheme is free from  
CMB distortion. To be compatible with the COBE-FIRAS CMB observation, we need to suppress the non-resonant conversion probability (\ref{eq:non-resonant-prob}) as
\bea
P_{\gamma\leftrightarrow a,\gamma^\prime}(\omega\sim \omega_{\rm CMB})\,\simeq\, \frac{1}{2} \frac{g^2_{a\gamma\gamma^\prime}}{g^2_{a\gamma^\prime\gamma^\prime}}
\lesssim 10^{-4},\eea
which requires\bea
\label{coupling_hierarchy}
\frac{g_{a\gamma\gamma^\prime}}{g_{a\gamma^\prime\gamma^\prime}} \lesssim 1.4\times 10^{-2}.\eea

The conversion probability $P_{\gamma\leftrightarrow \phi^+}$ given in (\ref{eq:prob_small_mixing}) and (\ref{eq:ConvRateLargeDarkMixing}) indicates that the ALP coupling ratio $g_{a\gamma\gamma^\prime}/g_{a\gamma^\prime\gamma^\prime}$ is a key parameter to determine the size of
the conversion rate at $\omega=\omega_{21}$.
 In Fig.~\ref{fig:Pres21cm}, we plot the contours of $P_{\gamma\leftrightarrow \phi^+}(\omega=\omega_{21})$ in the parameter space  $(m_a, g_{a\gamma\gamma^\prime}/g_{a\gamma^\prime\gamma^\prime})$ for $g_{a\gamma^\prime\gamma^\prime}=2\times 10^{-7} \, {\rm GeV}^{-1}$ and
${\cal B}_X=0.5$ $\mu$G. Our result shows that the conversion probability can be close to unity over a wide range of parameter space satisfying the observational constraints. This allows that our scheme can provide a viable explanation 
of the EDGES anomaly even with a small amount of DR ($\phi=a $ or $\gamma^\prime$) at the 21\,cm frequency range, e.g. 
$f_\phi^{\rm 21cm}\Delta N_{\rm eff}^{\phi}\sim 10^{-9}$. 

A key ingredient of our scheme is the existence of  a primordial background dark photon gauge field $\langle X_{\mu\nu}\rangle=(\vec E_X, \vec B_X)$. As was demonstrated in \cite{Choi:2018dqr}, even a large  $\langle X_{\mu\nu}\rangle$ close to the upper bound $\sim 1$ $\mu$G can be generated by an ultra-light ALP $\varphi$ which couples to the dark photon gauge field
through\bea\frac{1}{4} g_{\varphi\gamma^\prime \gamma^\prime}\varphi X_{\mu\nu}\tilde{X}^{\mu\nu}.\eea
In Appendix \ref{app:generation},
we provide an explicit scheme to generate the necessary
$\langle X_{\mu\nu}\rangle$ based on the results of \cite{Choi:2018dqr}.

Another key ingredient of our scheme to explain the EDGES anomaly is
the DR composed of $a$ or $\gamma^\prime$ with an energy density (in the 21\,cm frequency range) satisfying
\bea
\label{condition3}
f_\phi^{\rm 21cm}\Delta N_{\rm eff}^{\phi}\,\sim \,\frac{10^{-9}}{P_{\phi\rightarrow \gamma}}.
\eea
For the origin of such DR,
we can use the mechanisms proposed in \cite{Pospelov:2018kdh,Moroi:2018vci} utilizing the moduli or saxion decays into ALPs or the decays of another ALP
(constituting the dark matter) to  dark photons. Alternatively one can adopt the mechanism of \cite{Choi:1996vz} utilizing the decays of flaton to either ALPs or dark photons. As it is rather straightforward to apply the results of \cite{Pospelov:2018kdh,Moroi:2018vci} to our case, we do not provide a separate discussion on the generation of DR
satisfying the condition (\ref{condition3}).

As indicated by (\ref{bound_2}) and (\ref{coupling_hierarchy}), our scheme requires a hierarchical pattern of ALP couplings:
\bea
g_{a\gamma\gamma}, \, g_{a\gamma\gamma^\prime}\,\ll\, g_{a\gamma^\prime\gamma^\prime}.\eea 
There can be a variety of ways to achieve such hierarchical ALP couplings without causing a fine tuning problem.
One option is that the PQ-charged massive fermions in the underlying UV model are charged only under $U(1)_X$, while
there exist additional PQ-singlet massive fermions charged under both $U(1)_X$ and the SM hypercharge $U(1)_Y$ \cite{Kaneta:2016wvf,Kaneta:2017wfh}. Then the loops of
PQ-singlet massive fermions induces a kinetic mixing $\varepsilon ={\cal O}(e g_X/16\pi^2)={\cal O}(10^{-3}-10^{-2})$
 between $X_{\mu\nu}$ and $F_{\mu\nu}$, while the loops of PQ-charged massive fermions generates the ALP coupling $g_{a\gamma^\prime\gamma^\prime}$ without generating $g_{a\gamma\gamma}, \, g_{a\gamma\gamma^\prime}$ in the original field basis.
 Then,  rotating away the kinetic mixing by an appropriate field redefinition, we  obtain the ALP couplings 
\bea
g_{a\gamma\gamma}\sim \epsilon g_{a\gamma\gamma^\prime}\sim \epsilon^2g_{a\gamma^\prime\gamma^\prime}.\eea
Alternatively one may use the clockwork mechanism of \cite{Choi:2014rja, Choi:2015fiu, Kaplan:2015fuy}
 to generate a hierarchical pattern of ALP couplings as in  \cite{Higaki:2015jag,Farina:2016tgd,Agrawal:2017cmd}, which can give even a bigger hierarchy among the ALP couplings.
Note that the above pattern of ALP couplings is in good accordance with the astrophysical constraints  
\bea g_{a\gamma\gamma^\prime} < 5\times 10^{-10} {\rm GeV}^{-1},\quad 
 g_{a\gamma\gamma} < 5\times 10^{-12}  {\rm GeV}^{-1}
\eea
 which are deduced from the star cooling due to the plasmon decay
$\gamma_{\rm pl}\rightarrow a\gamma^\prime$ \cite{Choi:2018mvk}
and the absence of $\gamma$-ray burst associated with  SN1987A  \cite{Payez:2014xsa}.




\section{Conclusion}
\label{sec:con}
In this paper, we examined the ALP-photon-dark photon oscillations in background dark photon gauge field, while focusing on the resonant conversion between the photon and a dark radiation composed of ALPs and dark photons in the early Universe. We find that the corresponding conversion probability reveals an interesting spectral feature which allows strong
conversion at low frequency domain, but has negligible conversion at high frequencies above certain critical frequency  which is  determined by the ALP coupling to dark photon and the strength of background dark photon gauge field. We then utilize this feature to 
heat up  the 21\,cm photons without affecting the Cosmic Microwave Background, which may explain the recent tentative observation by the EDGES experiment of an anomalously strong absorption signal of 21\,cm photons.
We find that our scheme can explain the EDGES anomaly over a wide range of parameter space, while satisfying the observational constraints and also without causing a naturalness problem.


\begin{acknowledgments}
This work was supported by IBS under the project code IBS-R018-D1.
We thank S. Lee and C. S. Shin for helpful discussions.
\end{acknowledgments}

\appendix

\section{A brief review of resonant conversion between photon and dark radiation}
\label{app:Overview}

Here we briefly review the cosmological resonant conversion between the photon $\gamma$ and a light hidden sector particle $\phi$ such as  ALP or dark photon, which
is a straightforward generalization of the well known Mikheyev-Smirnov-Wolfenstein (MSW) effect in neutrino physics  \cite{Wolfenstein:1977ue,Mikheev:1986gs,Mikheev:1986wj}.
For this, let us consider the effective mass-square matrix of $\gamma$ and $\phi$ in generic time-dependent environment:
\bea
\label{evo}
\mathcal{M}^2 = \left[
\begin{tabular}{cc}
$m_{\gamma}^2$ & $m_{\rm mix}^2$ \\
$m_{\rm mix}^2$ & $m_{\rm \phi}^2$
\end{tabular}
\right] \, ,
\eea
where $m_{\gamma}$ is the effective photon mass induced by the scattering off the ambient medium, and $m_{\rm mix}$ describes the mixing induced by an appropriate coupling of $\phi$ to the photon. 
The evolution equation for relativistic propagation of $\gamma$ and $\phi$ is given by
\bea
\left[i\frac{d}{dt} - \frac{1}{2\omega}\mathcal{M}^2\right] \left(
\begin{array}{c}
\gamma \\
\phi
\end{array}
\right) = 0 \, ,
\eea
where $\omega$ is the energy of the state. Here
we are interested in the case that ${\cal M}^2$ varies in time due to the expansion of the early universe.

To proceed, one may rewrite the above evolution equation in the instantaneous mass eigenbasis $(\psi_+,\psi_-)$ as 
\bea
\label{evol}
\left[i\frac{d}{dt} - 
\left(
\begin{array}{cc}
\frac{1}{2\omega}(m_\gamma^2+m_\phi^2)+\Delta_{\rm osc} & -i\frac{d\chi}{dt} \\
i\frac{d\chi}{dt} & \frac{1}{2\omega}(m_\gamma^2+m_\phi^2)-\Delta_{\rm osc}
\end{array}
\right)
\right] \left(
\begin{array}{c}
\psi_+ \\
\psi_-
\end{array}
\right) = 0 \, .
\eea
where 
\bea
\left(
\begin{array}{c}
\psi_+ \\ \psi_-
\end{array}
\right) = \left(
\begin{array}{cc}
\cos\chi & -\sin\chi \\
\sin\chi & \cos\chi
\end{array}
\right)
\left(\begin{array}{c}
\gamma \\ \phi
\end{array}
\right) \, ,
\eea
and the instantaneous mixing angle and oscillation frequency are given by
\bea
\tan2\chi =  \frac{2m_{\rm mix}^2}{m_\gamma^2 - m_\phi^2}, \quad
\Delta_{\rm osc} = \frac{\sqrt{\left(m_{\gamma}^2-m_\phi^2\right)^2 + 4 m_{\rm mix}^4}}{4\omega}.\eea
The above evolution equation indicates that the transition between $\psi_+$ and $\psi_-$ is determined by the adiabaticity  parameter
\bea
\gamma_{\rm ad}=\frac{\Delta_{\rm osc}}{|d\chi/dt|},\eea
which is large in the adiabatic limit $|d\chi/dt|\ll \Delta_{\rm osc}$. 
In our case, the time variance of the mixing angle $\chi$ originates from the expansion of the universe. Then, in the small mixing regime with
$\sin\chi \simeq |m_{\rm mix}^2/(m_\phi^2-m_\gamma^2)|\ll 1$, we have $d\chi/dt ={\cal O}(H\sin\chi)$, where $H$ is the Hubble expansion rate, while
$d\chi/dt ={\cal O}(H\times {\rm Max}(m_\phi^2,m_\gamma^2)/m_{\rm mix}^2)$ in the large mixing regime with $|m_{\rm mix}^2/(m_\phi^2-m_\gamma^2)|\gg 1.$

Here we are concerned with the conversion of an initial photon (or $\phi$) at $t_i$ to the hidden sector particle $\phi$ (or photon) in the final state at $t_f$.
We can then consider two distinctive cases.  The first case is that there occurs a sign flip of $m_\gamma^2-m_\phi^2$ during the evolution, e.g. $m_\gamma^2(t_i)> m_\phi^2(t_i)$, but $m_\gamma^2(t_f) < m_\phi^2(t_f)$, so there exists
a resonance point 
where \bea m^2_\gamma(t_{\rm res})=m^2_\phi(t_{\rm res})  \quad (t_i < t_{\rm res}<t_f).\eea
The other case is that $m_\gamma^2(t)> m_\phi^2(t)$ over the entire evolution from $t_i$ to $t_f$, so there is no resonant point in between. Note that $0<\chi_i< \frac{\pi}{4}$ and $\frac{\pi}{4}<\chi_f<\frac{\pi}{2}$, and therefore $\cos 2\chi_f\cos 2\chi_i <0$ for the first case, while $ 0<\chi_i, \chi_f <\frac{\pi}{4}$ for the other case, yielding $\cos 2\chi_f\cos 2\chi_i >0$.

If the relevant time intervals such as $t_f-t_{\rm res}$ and $t_{\rm res}-t_i$ are much longer than the oscillation length $\Delta_{\rm osc}^{-1}$, one can take an average over the production and detection points.
In the adiabatic limit that $d\chi/dt$ in the evolution equation (\ref{evol}) can be ignored, one easily finds
the conversion probability averaged over the production and detection points is given by 
\bea
P_{\gamma\leftrightarrow \phi} & = & \frac{1}{2}-\frac{1}{2}\cos 2\chi_f\cos 2\chi_i ,
\eea
where $\chi_i$ and $\chi_f$ denote the mixing angles at the production and detection points, respectively. One can now include the effects of nonzero $d\chi/dt$ in the evolution.
 In our case,
 $\gamma_{\rm ad}=\Delta_{\rm osc}/|d\chi/dt|={\cal O}(\Delta_{\rm osc}/H\sin\chi)\gg 1$   {\it except} near the resonance point.
Then  the effects of time-varying mixing angle can be included in the transition probability as \cite{Parke:1986jy}
\bea
\label{prob1}
P_{\gamma\leftrightarrow \phi} & = & \frac{1}{2}-\left(\frac{1}{2}-p_{\rm res}\right)\cos 2\chi_f\cos 2\chi_i +{\cal O}\left(\frac{H^2\sin\chi^2}{\Delta_{\rm osc}^2}\right)\, ,
\eea
where $p_{\rm res}$ is the probability for the level crossing between $\psi_+$ and $\psi_-$ at the resonance point,
and the last term is an order of magnitude estimate of the transitions between $\psi_+$ and $\psi_-$ that occur outside the resonance region.  
In case that there is no resonance point during the evolution, one can simply set $p_{\rm res}=0$ to get the corresponding conversion probability.

If the resonance time interval is short enough that  $m_\gamma^2 -m_\phi^2$ can be approximated as a linear function of time, which holds for our case,
one can use the Landau-Zener result to find \cite{Landau,Zener:1932ws}
\bea
\label{prob2}
p_{\rm res} =\exp\left(-\frac{\pi {\gamma_{\rm ad}} (t_{\rm res})}{2}\right)\,
\simeq\,  \left.\exp \left( - r\frac{\pi m_{\rm mix}^4}{\omega  m_\phi^2}\right) \right|_{t=t_{\rm res}} \, , 
\label{eq:LandauZener}
\eea
where 
\bea
r^{-1}= \left|\frac{d \ln (m_{\gamma}^2(t)/m_\phi^2(t))}{dt}\right|_{t=t_{\rm res}} \,=\, {\cal O}(1-10)\times H(t_{\rm res})\, .
\eea
In case that $\gamma_{\rm ad}(t_{\rm res}) \gg 1$ and therefore the evolution is adiabatic even at the resonance point,
the level crossing probability $p_{\rm res}$ is exponentially small, which results in a large transition probability as
\bea
P_{\gamma\leftrightarrow\phi} = \frac{1}{2}+\frac{1}{2}|\cos 2\chi_f\cos 2\chi_i| +  {\cal O}\left(\frac{H^2\sin\chi^2}{\Delta_{\rm osc}^2}, \exp\left(-\frac{\pi\gamma_{\rm ad}(t_{\rm res})}{2}\right)\right).\eea
 Note that $\cos 2\chi_f\cos 2\chi_i <0$ in this case.
On the other hand, if $\gamma_{\rm ad}(t_{\rm res})\ll 1$, which means that adiabaticity is abruptly violated at the resonance point,  the level crossing probability $p_{\rm res}$ is close to the unity and then
the transition probability can be approximated as
\bea
P_{\gamma\leftrightarrow\phi} = \frac{1}{2}\left(1-|\cos 2\chi_f\cos 2\chi_i|\right) +\frac{\pi}{2} 
|\cos 2\chi_f\cos 2\chi_i|\gamma_{\rm ad}(t_{\rm res})+{\cal O}\left(\frac{H^2\sin\chi^2}{\Delta_{\rm osc}^2}\right), \eea
In fact, in this case  the initial and final mixing angles have small values  as $\chi_i^2, \chi_f^2 <\gamma_{\rm as}(t_{\rm res})\ll 1$, so the conversion probability can be approximated by the following simple form:
\bea
P_{\gamma\leftrightarrow\phi} \,\simeq\, \frac{\pi\gamma_{\rm ad}(t_{\rm res})}{2}
\,=\,
\left. r\frac{\pi m_{\rm mix}^4}{\omega m_\phi^2} \right|_{t=t_{\rm res}} , \label{eq:nonResConv}
\eea
where 
$r^{-1}= \left|d\ln (m_{\gamma}^2/m_\phi^2)/{dt}\right|_{t=t_{\rm res}} \,=\, {\cal O}(1-10)\times H(t_{\rm res})$.

 \section{Generation of background dark photon gauge field}
 \label{app:generation}

 To complete our scheme, we need an explanation for the origin of the  primordial background dark photon gauge field $\langle X_{\mu\nu}\rangle$,
Here we discuss an explicit scheme to generate the required $\langle X_{\mu\nu}\rangle=(\vec E_X, \vec B_X)$, which is based on the mechanism of \cite{Choi:2018dqr}. 
For this, we introduce an additional ultra-light ALP $\varphi$ which couples to the massless dark photon gauge field $X_\mu$ as
\begin{equation}
\label{coupling_gen}
\frac{1}{4} g_{\varphi\gamma^\prime \gamma^\prime}\varphi X_{\mu\nu}\tilde{X}^{\mu\nu}.
\end{equation}
Around the time $t_{\rm osc}$ when the Hubble expansion rate $H(t_{\rm osc})\sim m_{\varphi}$, the ultra-light ALP $\varphi$ begins to oscillate as
\bea
\varphi \simeq \varphi_i \left(\frac{R(t)}{R(t_{\rm osc})}\right)^{-3/2}\cos\left(m_{\varphi}(t-t_{\rm osc})\right),\eea
where $R(t)$ is the scale factor of the expanding Universe with the spacetime metric $ds^2=dt^2-R^2(t)d\vec x^2$.
The oscillating $\varphi$ causes a tachyonic instability of $X_\mu$ through the ALP coupling (\ref{coupling_gen})
 for the
 wave numbers $k\sim g_{\varphi\gamma^\prime\gamma^\prime}\dot\varphi$, thereby amplifies the quantum fluctuation of $X_\mu$
 to a stochastic classical background field.  For efficient amplification,  $g_{\varphi\gamma'\gamma'}\varphi_i$
 needs to be large enough to overcome the dilution
 by the Hubble expansion, but not too large to avoid a too strong friction which would forbid the oscillatory motion of $\varphi$. It was found in \cite{Kitajima:2017peg,Choi:2018dqr,Agrawal:2018vin} that this can be achieved  when \bea
 g_{\varphi\gamma'\gamma'}\varphi_i= {\cal O}(10-100),\eea which  will be assumed here.
Then we can use the results of \cite{Choi:2018dqr} to find that the amplified dark photon gauge field today
is determined  by the initial ALP misalignment as
\bea
\label{result1}
|\vec E_X(t_0)|\sim |\vec B_X(t_0)|\sim 0.3 \,\mu{\rm G}\left(\frac{\varphi_i}{10^{17}\, {\rm GeV}}\right), \eea
while the red-shift $z_*$ at the time of amplification (production) and the coherent length $\lambda$ of the produced fields today are determined by the ALP mass as\bea
\label{result2}
z_* \,\sim\,  10^4 \bigg(\frac{m_\varphi}{10^{-25}\text{eV}}\bigg)^{1/2}, \quad
\lambda \,\sim\, 3 \, {\rm Mpc}\bigg(\frac{m_\varphi}{10^{-25}\text{eV}}\bigg)^{-1/2}.\eea
This process leaves also a coherently oscillating ALP dark matter whose mass density is given by
\bea
\label{result3}
\frac{\Omega_\varphi h^2}{0.12}\simeq 10^{-3}\,\bigg(\frac{m_\varphi}{10^{-25}\text{eV}}\bigg)^{1/2}
\left(\frac{\varphi_i}{10^{17}\,{\rm GeV}}\right)^2.\eea

The energy density of the produced $\vec E_X$ and $\vec B_X$ is red-shifted like a radiation energy density, so is
bounded as $\Delta N^X_{\text{eff}}\leq 0.3$,  where  
\bea
\Delta N^X_{\text{eff}}\simeq 0.4 \left(\frac{\sqrt{\vec E_X^2(t_0) +\vec B_X^2(t_0)}}{1\,\mu\text{G}}\right)^2,\eea
This implies that the background dark photon gauge field combination
$\vec{\cal B}_X =\vec B_X-\hat k (\hat k\cdot\vec B_X) -\hat k\times \vec E_X$ which is
relevant for the ALP-photon-dark photon oscillation 
is roughly bounded  as \bea
{\cal B}_X\lesssim 1 \, \mu{\rm G}.\eea
With the results  (\ref{result1}), (\ref{result2}) and (\ref{result3}), we can choose the ALP parameters $\varphi_i={\cal O}(10^{17})$ GeV,
$m_\varphi={\cal O}(10^{-25})$ eV and $g_{\varphi\gamma^\prime\gamma^\prime}\varphi_i={\cal O}(10-10^2)$ to generate
${\cal B}_X={\cal O}(0.1-1) \,\mu{\rm G}$ well before the recombination, e.g. $z \gtrsim 10^4$,
together with  $\Omega_\varphi$ which is small enough to satisfy the observational bounds summarized in \cite{Grin:2019mub}.
Note that this production of ${\cal B}_X$ driven by $\varphi$ is not affected by the existence of the other ALP $a$. Although  $g_{a\gamma^\prime\gamma^\prime}\gg
g_{\varphi\gamma^\prime\gamma^\prime}$ to explain the EDGES anomaly,
 as long as $m_a/m_{\varphi}$ is large enough, which can be as large as $10^{16}$ in our case, the heavier ALP $a$
is safely decoupled from the slow dynamics of $\varphi$ producing ${\cal B}_X$ at late time with the Hubble expansion rate $H(t) < m_{\varphi}$.

 \section{Observational constraints on the ALP to photon conversion scenario}
 \label{app:ALPrevision}

\begin{figure}[t]
\centering
\includegraphics[width=0.55\textwidth]{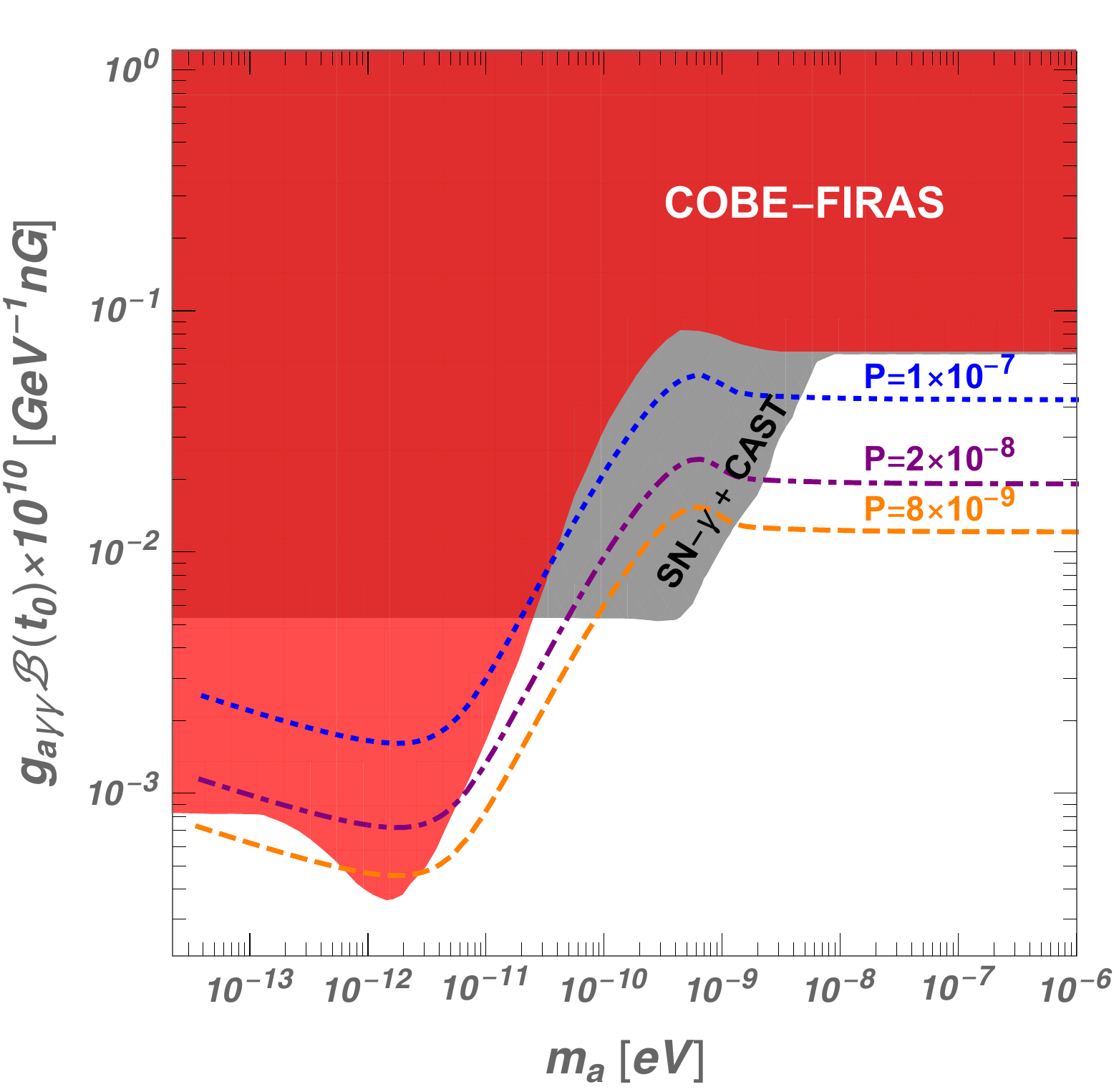}
\caption{Parameter region of the ALP scenario of \cite{Moroi:2018vci} excluded by the observational constraints.  Dotted curves are the contours for three different values of the conversion probability: $P_{a\rightarrow \gamma}=10^{-7},\, 2\times 10^{-8}, 8\times 10^{-9}.$ }
\label{fig:ALP}
\end{figure}

Here we revisit the ALP scenario of  \cite{Moroi:2018vci} and update the observational constraints on the model parameters.
Our results are summarized in Fig.~\ref{fig:ALP}.

The red-colored region in Fig.~\ref{fig:ALP} which was derived in \cite{Moroi:2018vci} is excluded as it results in a too large distortion of the CMB spectrum probed by the COBE-FIRAS.  The gray-colored region is excluded by combining
the constraint on
$g_{a\gamma\gamma}$
from the absence of $\gamma$-ray burst associated with SN1987A  \cite{Payez:2014xsa} and
the recently derived 
 upper bound 
 $B_0 \lesssim 0.1 \, {\rm nG}$ on the primordial magnetic field to avoid an overheating of baryons which would wash away the EDGES signal \cite{Minoda:2018gxj}.
One may consider also the constraint from blackhole supperradiance, which is known to exclude the ALP mass range
$7\times 10^{-14} {\rm eV} < m_a < 2\times 10^{-11}{\rm eV}$ at 95 \% confidence level 
if the ALP quartic coupling is weak enough \cite{Stott:2018opm}.
However this does not apply for the present case as the axion decay constant $f_a$ suggested by  the size of $g_{a\gamma\gamma}$ implies that the corresponding axion quartic coupling
$\lambda\sim m_a^2/f_a^2$ is large enough to invalidate the  blackhole supperradiance argument.

In Fig.~\ref{fig:ALP}, the dotted lines show the parameter region yielding the minimal 
conversion probability $P_{a \rightarrow \gamma}={\cal  O}(10^{-8}-10^{-7})$ which is required to explain the EDGES anomaly.
 We then  find that only a tiny parameter region of $(m_a, g_{a\gamma\gamma})$ can provide a viable explanation for the EDGES anomaly, but only when both $B_0$ and the ALP number density in the 21\,cm frequency region
 nearly saturate their upper bounds.

\end{document}